\documentclass[pdflatex,sn-mathphys-num]{sn-jnl}

\usepackage{graphicx}%
\usepackage{multirow}%
\usepackage{amsmath,amssymb,amsfonts}%
\usepackage{amsthm}%
\usepackage{mathrsfs}%
\usepackage[title]{appendix}%
\usepackage{xcolor}%
\usepackage{textcomp}%
\usepackage{manyfoot}%
\usepackage{booktabs}%
\usepackage{algorithm}%
\usepackage{algorithmicx}%
\usepackage{algpseudocode}%
\usepackage{listings}%

\theoremstyle{thmstyleone}%
\theoremstyle{thmstyletwo}%

\theoremstyle{thmstylethree}%

\usepackage{amsmath,amssymb,bm}
\usepackage{tikz}
\usetikzlibrary{arrows.meta,positioning,calc,shapes.geometric,fit,backgrounds}
\tikzset{
  >=Latex,
  state/.style={circle,draw,thick,minimum size=9.5mm,align=center},
  iface/.style={rectangle,draw,thick,rounded corners,
                minimum width=20mm,minimum height=7mm,align=center},
  outbox/.style={rectangle,draw,thick,rounded corners,
                 minimum width=24mm,minimum height=8mm,align=center},
  startdot/.style={circle,fill=black,inner sep=1.5pt,outer sep=0pt},
  sum/.style={circle,draw,thick,double,minimum size=9.5mm,inner sep=0pt,font=\large},
  lab/.style={font=\small,inner sep=1pt,outer sep=1pt},
  trans/.style={-Latex,thick},
  soft/.style={-Latex,dashed,thick},
  note/.style={font=\footnotesize,align=left}
}

\begin{document}


\title{Representation-induced superposition breakdown in linear physics}


\author[1]{\fnm{Michael} \sur{Mazilu}}\email{Michael.Mazilu@st-andrews.ac.uk}

\author[2]{\fnm{Andriejus} \sur{Dem\v{c}enko}}\email{Andriejus.Demcenko@glasgow.ac.uk}

\affil[1]{\orgdiv{SUPA, School of Physics and Astronomy}, \orgname{University of St. Andrews}, \orgaddress{\street{North Haugh}, \city{St. Andrews}, \postcode{KY16 9SS}, \state{Scotland}, \country{United Kingdom}}}

\affil[2]{\orgdiv{Division of Biomedical Engineering, School of Engineering}, \orgname{The University of Glasgow}, \orgaddress{\street{University Avenue}, \city{Glasgow}, \postcode{G12 8LT}, \state{Scotland}, \country{United Kingdom}}}

\abstract{The superposition principle is fundamental to linear wave systems, ensuring that their physical behaviour is independent of the chosen basis representation. While this principle underpins many analytical techniques, including modal decompositions and scattering formulations, we show that superposition expansion can fail in multilayered media when fields are expressed as infinite series of evanescent and inhomogeneous waves. Using the Airy formula and the scattering-matrix formalism, we identify conditions under which the superposition of partial waves diverges, particularly in systems with three or more interfaces. This divergence occurs because evanescent wave components cannot be normalised within the conventional basis and is not a numerical artefact. To address this, we introduce power flux modes corresponding to orthonormal basis wave solutions that preserve energy conservation in scattering events and consequently restore convergence. We prove that in the flux‑orthonormal basis, interface scattering is unitary and propagation eigenvalues are bounded ($\le 1$), guaranteeing convergence.
Our approach generalises to scalar, electromagnetic, and elastic wave systems, providing a robust framework for eliminating evanescent mode divergence without regularisation or renormalisation.
}




\maketitle

\section{Introduction}\label{sec1}

The superposition principle is foundational to linear wave theory: solutions to linear wave equations are representable as linear combinations of basis functions, a perspective that underpins standard analytical tools from modal decompositions to waveguide theory and scattering formulations \cite{Jackson1999,LandauLifshitz1977}. Beyond canonical transfer- and scattering-matrix treatments, related multiple-scattering frameworks
\cite{bohren1983absorption,foldy1945multiple} and periodic-media analyses in photonic crystals \cite{joannopoulos2008photonic} highlight how representation choices (e.g. basis and normalisation) govern both convergence and physical interpretability.
In optics, multilayer interference is classically described by the Airy formulation and by transfer/scattering-matrix methods, which sum infinite series of partial waves associated with successive reflections and transmissions \cite{BornWolf1999,Yeh1988,KoSambles1988}. These approaches are extremely successful for stacks that only support propagating modes, and their stability can be further enhanced by scattering‑matrix schemes that are well behaved near total internal reflection \cite{KoSambles1988,Berreman1972}. 

However, in stratified systems that support evanescent (imaginary wavevector normal to the interface) or inhomogeneous waves (complex wavevector normal to the interface), the usual Airy-type infinite series can fail to converge when fields are expanded in conventional plane‑wave amplitudes \cite{Yeh1988,KoSambles1988}. This divergence is not a failure of linearity per se, but a consequence of expressing the solution in a basis that includes components that cannot be normalised with respect to conserved energy flux; in such cases, the multiple‑scattering series diverges for specific parameter ranges.
The phenomenon is physically tied to the role of evanescent fields that decay spatially and do not carry a net power flux across layers and to the spectral properties of the event‑scattering operator in multilayers \cite{NovotnyHecht2012}. Analogous convergence pathologies are long recognised in multiple-scattering series,
where expansions for coherent fields or effective parameters can fail or require resummation beyond the leading orders
\cite{foldy1945multiple}.

To resolve this, we restrict the solution space to \emph{power‑flux eigenmodes}: orthonormal modes defined with respect to the normal component of the Poynting vector, ensuring that each retained mode contributes a finite and physically meaningful amount of energy and that the corresponding series converges without ad hoc regularisation \cite{Marcuse1991}. This construction preserves unitarity at interfaces and bounds the eigenvalues associated with intra‑layer propagation, restoring convergence of the Airy‑type expansion while remaining compatible with standard multilayer formalisms \cite{KoSambles1988,Berreman1972}. 

Beyond optics, analogous issues and remedies arise across linear wave physics, with similar convergence and normalisation challenges appearing in elastic‑wave multiple scattering and layered‑media formulations \cite{foldy1945multiple,joannopoulos2008photonic}.
The same considerations apply to elastic waves in layered solids, where modal completeness and power‑based orthogonality are central (see also the elastic analogues discussed in Appendix C), and to quantum systems in which probability current plays the role of power flux and unitarity constrains scattering \cite{LandauLifshitz1977,Sprague2024UnitarityQI,Kucera2025CurrentSMatrix}.
While divergent wave expansions are often resolved by resummation or renormalisation techniques (e.g. Cesàro‑type or related summation approaches used historically in multiple‑scattering theory) \cite{foldy1945multiple,anderson1958absence,Nellambakam2023,Takatsu.Sebilleau.2022},
our strategy is different: by constructing an orthonormal basis with respect to the \emph{power‑flux} (Poynting) measure, we restore unitarity and convergence at the representation level without external summation prescriptions \cite{Marcuse1991,KoSambles1988,Wojcik2021}.
Indeed, recent advances in nanophotonics and quantum photonics show that the evanescent-mode governs the scattering behaviour in photonic devices \cite{Wojcik2021,He2024,ChouChau2025,Zhu2025}, while quantum‑information platforms increasingly rely on multi-mode photonic scattering models \cite{Wang2025,AbuGhanem2026,Kolarovszki2025}.
Furthermore, in quantum computing and information, unitarity is the fundamental constraint when considering scattering \cite{Sprague2024UnitarityQI} and photonic structures \cite{Briceno2024PhotonicQC}. Quantum transport analyses similarly show that only flux carrying modes form a unitary S‑matrix \cite{Kucera2025CurrentSMatrix}.

In what follows, we: (i) revisit the Airy formalism and identify the exact conditions under which divergence occurs in three‑interface stacks; (ii) develop the power‑flux eigenmode basis and its interface/propagation operators; and (iii) demonstrate numerically and analytically that this basis eliminates evanescent‑mode divergence while preserving the physical content of the multilayer problem. 

\subsection{Airy formula and its divergence}

\begin{figure}
    \centering    \includegraphics[width=0.95\linewidth]{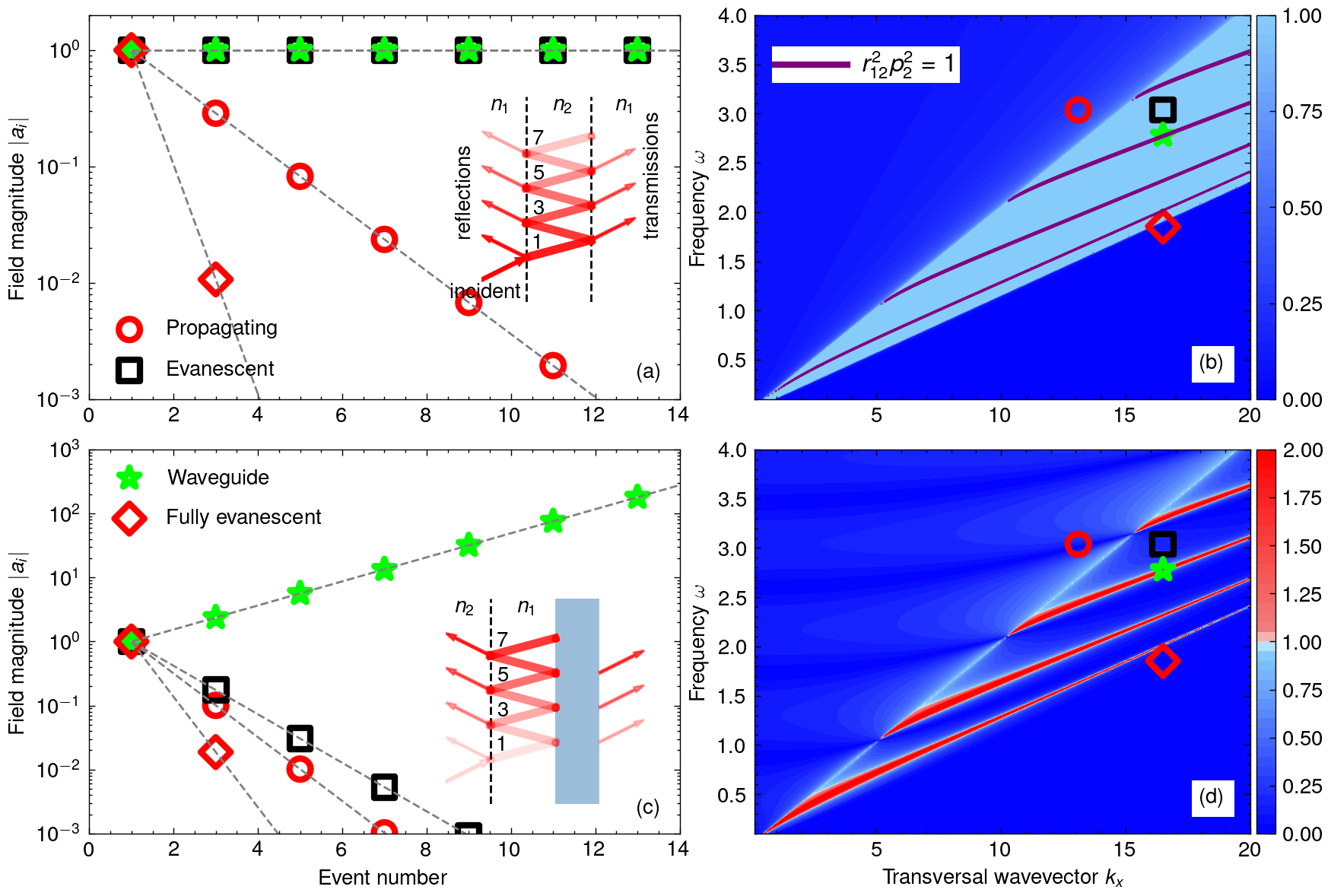}
    \caption{Internal field magnitude: (a) Single layer with two interfaces and (c) two layer system calculated using the same single layer formalism where the reflection and transmission coefficients of the second interface are replaced ones calculated from the second (gray) layer. (b,d) Magnitude of the round-trip coefficient $|r_{12}^2p_2^2|$ corresponding respectively to the structures considered in figures (a) and (c). }
    \label{fig:1}
\end{figure}

The Airy formula \cite{Yeh1988} describes the transmission and reflection of light through plane parallel layers. 
Conceptually, it breaks down the interaction between the wave and the layers into a sequence of ordered events, each corresponding to physical phenomena such as reflection, transmission, or propagation. 
A typical example of this approach is to consider the overall transmission $T_{13}$ through a single layer $(1|2|3)$ as the sum of all the individual partial waves resulting from each event (see inset Figure \ref{fig:1}a) 
\begin{align}
    T_{13} &= t_{12}t_{23}p_2 \left(1 + r_{21}r_{23}p_2^2 + (r_{21}r_{23}p_2^2)^2 + \ldots \right) =\frac{t_{12}t_{23}p_2}{1 - r_{21}r_{23}p_2^2} \\
    R_{13} & = r_{12} + t_{12}\, p^2_{2} \, r_{23} t_{21}  \sum_{n=0}^{\infty} \big(  r_{21}r_{23}p_2^2\big)^n = r_{12} + \frac{t_{12}\, p^2_{2} \, r_{23} t_{21}}{1 - r_{21}r_{23}p_2^2}
\end{align}
where $r_{ij}= \frac{k_{zi} - k_{zj}}{k_{zi} + k_{zj}}$ and $t_{ij}= \frac{2 k_{zi}}{k_{zi} + k_{zj}}$ are the Fresnel coefficients of the interface for reflection and transmission with $k_{zj}$ the $z$ components (direction normal to the interface) of the wavevectors in the medium $j$ (or $i$). 
The propagation coefficient $p_j=\exp(\imath\, k_{zj}h_j)$ takes into account the optical path length of the layer with thickness $h_2$, where $\imath=\sqrt{-1}$. 
Similarly, a second Airy formula describes the overall reflection coefficient $R_{13}$ as the sum of all individual Fresnel reflections and transmission events, leading to a final partial wave propagating in the reflection direction. 
Both formulas are based on the convergences of the sum of a geometric series conditioned by $|r_{21}r_{23}p_2^2| < 1$ represented as contour plots in Figs. \ref{fig:1}b,d. 
This geometric-series structure mirrors standard multireflection and multipole expansions used for particle and obstacle scattering
\cite{bohren1983absorption}.
The Airy formula is always convergent for a single-layer system with two interfaces, irrespective of the refractive indices and the angle at which light is incident (see Figs. \ref{fig:1}a,b). 

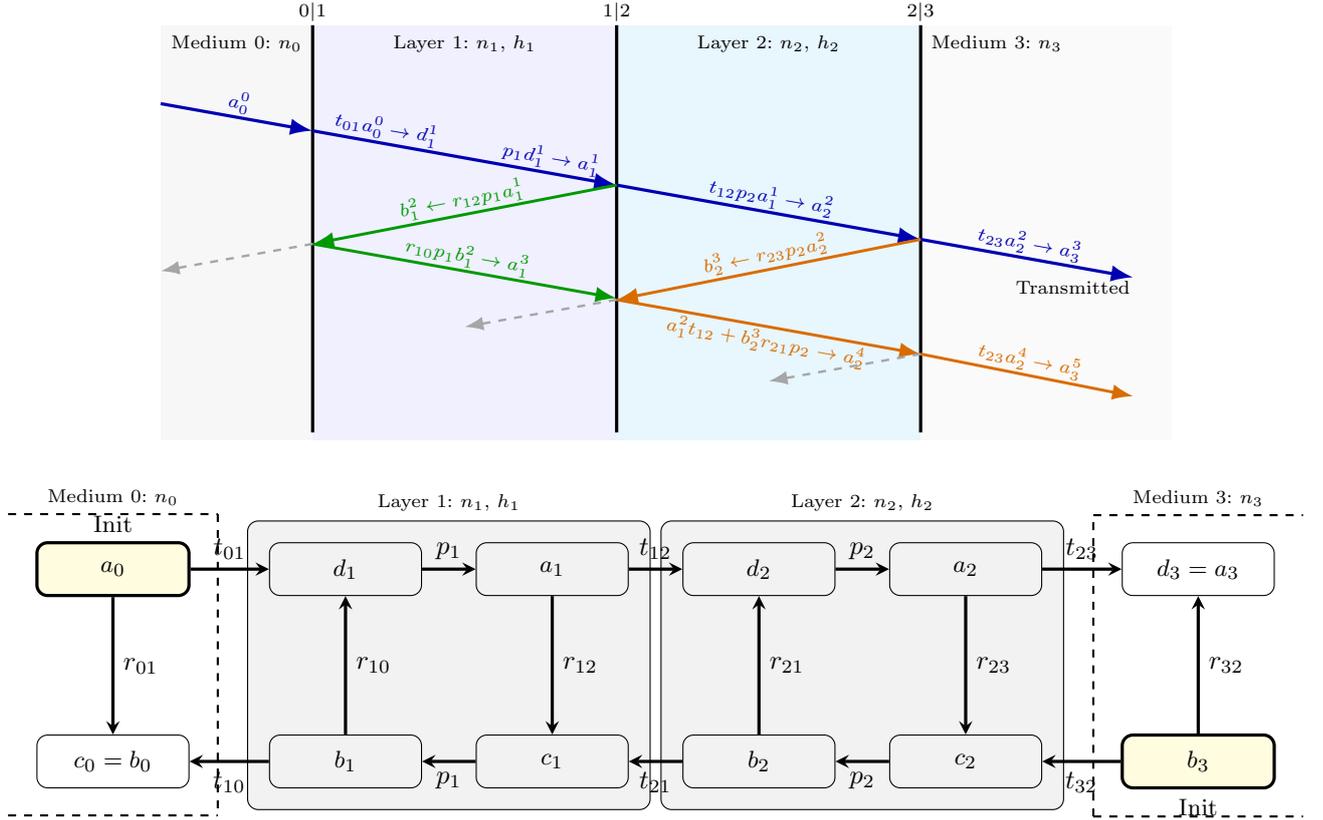
\begin{figure}
    \begin{center}

\begin{tikzpicture}[x=1cm,y=1cm,>=Latex, on grid]
\tikzset{
  iface/.style={very thick},
  layerbox/.style={draw, rounded corners, fill=gray!10, inner sep=6pt},
  ray0/.style={-Latex, line width=1.2pt, blue!70!black},
  ray1/.style={-Latex, line width=1.1pt, green!60!black},
  ray2/.style={-Latex, line width=1.1pt, orange!85!black},
  back/.style={-Latex, line width=0.9pt, dashed, gray!70},
  lbl/.style={font=\footnotesize, inner sep=1pt}
}

\def\xI{0}    \def\xJ{4}    \def\xK{8}    \def\xL{10.8} 

\def\sr{-0.18} \def\sl{ 0.18} \def\srr{-0.14} \def\sll{ 0.14} \def\srr{-0.18} \def\sll{ 0.18} 

\coordinate (S0)  at (-2,0.8);                           \coordinate (S0)  at (-2,{-\sr*2});                           \coordinate (I01) at ({\xI},  {0});                      \coordinate (I12) at ({\xJ},  {0 + \sr*(\xJ-\xI)});      \coordinate (I23) at ({\xK},  {0 + \srr*(\xK-\xI)});      \coordinate (OUT) at ({\xL},  {0 + \sr*(\xL-\xI)});      

\begin{scope}[on background layer]
  \fill[gray!06] (-2,-4.1) rectangle ({\xI}, 1.4);          \fill[blue!6]  ({\xI},-4.1) rectangle ({\xJ}, 1.4);       \fill[cyan!8]  ({\xJ},-4.1) rectangle ({\xK}, 1.4);       \fill[gray!04] ({\xK},-4.1) rectangle ({\xL+0.5}, 1.4);   \end{scope}

\draw[iface] ({\xI},-4.)--({\xI},1.4) node[above, lbl] {$0|1$};
\draw[iface] ({\xJ},-4.)--({\xJ},1.4) node[above, lbl] {$1|2$};
\draw[iface] ({\xK},-4.)--({\xK},1.4) node[above, lbl] {$2|3$};

\node[lbl, anchor=north] at ({(\xI+\xJ)/2}, 1.3) {Layer 1: $n_1$, $h_1$};
\node[lbl, anchor=north] at ({(\xJ+\xK)/2}, 1.3) {Layer 2: $n_2$, $h_2$};
\node[lbl, anchor=north] at ({\xI-1.0}, 1.3) {Medium 0: $n_0$};
\node[lbl, anchor=north] at ({\xK+1.0}, 1.3) {Medium 3: $n_3$};

\draw[ray0] (S0)--(I01) node[midway, above, sloped, lbl] {$a^0_0$};
\draw[ray0] (I01)--(I12) node[midway, above, sloped, lbl] {$t_{01}a^0_0\rightarrow d^1_1$ \hspace{22pt}$p_1d^1_1\rightarrow a^1_1$};
\draw[ray0] (I12)--(I23) node[midway, above, sloped, lbl] {$t_{12}p_2a^1_1\rightarrow a^2_2$};
\draw[ray0] (I23)--(OUT) node[midway, above, sloped, lbl] {$t_{23}a^2_2\rightarrow a^3_3$};

\def\eps{0.20} \coordinate (I12a) at ({\xJ}, {\sr*(\xJ-\xI) + \eps});
\coordinate (L01a) at ({\xI}, {\sr*(\xJ-\xI) + \eps + \sl*(\xI-\xJ)});
\coordinate (I12b) at ({\xJ}, {\sr*(\xJ-\xI) + \eps + \sl*(\xI-\xJ) + \sr*(\xJ-\xI)});
\coordinate (I23a) at ({\xK}, {\sr*(\xJ-\xI) + \eps + \sl*(\xI-\xJ) + \sr*(\xK-\xI)});
\coordinate (OUTa) at ({\xL}, {\sr*(\xJ-\xI) + \eps + \sl*(\xI-\xJ) + \sr*(\xL-\xI)});

\def\del{-0.22} \coordinate (I23b) at ({\xK}, {\sr*(\xK-\xI) + \del});
\coordinate (I12c) at ({\xJ}, {\sr*(\xK-\xI) + \del + \sl*(\xJ-\xK)});
\coordinate (I23c) at ({\xK}, {\sr*(\xK-\xI) + \del + \sl*(\xJ-\xK) + \sr*(\xK-\xJ)});
\coordinate (OUTb) at ({\xL}, {\sr*(\xK-\xI) + \del + \sl*(\xJ-\xK) + \sr*(\xL-\xK)});

\draw[ray1] (I12) -- ({\xI}, {\sr*(\xJ-\xI) + \sl*(\xI-\xJ) - 0.06}) node[midway, above, sloped, lbl] {$b^2_1\leftarrow r_{12}p_1a^1_1$};
\draw[ray1] ({\xI}, {\sr*(\xJ-\xI) + \sl*(\xI-\xJ) - 0.06})
           -- ({\xJ}, {\sr*(\xJ-\xI) + \sl*(\xI-\xJ) - 0.06 + \sr*(\xJ-\xI)}) node[midway, above, sloped, lbl] {$r_{10}p_1b^2_1\rightarrow a^3_1$};

\draw[ray2] (I23) -- ({\xJ}, {\srr*(\xK-\xI) + \sll*(\xJ-\xK) - 0.08}) node[midway, above, sloped, lbl] {$b^3_2\leftarrow r_{23}p_2 a^2_2$};
\draw[ray2] ({\xJ}, {\srr*(\xK-\xI) + \sll*(\xJ-\xK) - 0.08})
           -- ({\xK}, {\srr*(\xK-\xI) + \sll*(\xJ-\xK) - 0.08 + \srr*(\xK-\xJ)}) node[midway, below, sloped, lbl] {$a^2_1t_{12}+b^3_2r_{21}p_2\rightarrow a^4_2$};

\draw[ray2] ({\xK}, {\srr*(\xK-\xI) + \sll*(\xJ-\xK) - 0.08 + \srr*(\xK-\xJ)})
    -- ({\xL}, {\srr*(\xK-\xI) + \sll*(\xJ-\xK)  + 0.08 + \srr*(\xK-\xJ)+ \srr*(\xK-\xJ)})  node[midway, above, sloped, lbl] {$t_{23}a^4_2\rightarrow a^5_3$};
    
\draw[back]  ({\xI}, {\sr*(\xJ-\xI) + \sl*(\xI-\xJ) - 0.06}) 
  --  (-2, {\sr*(\xJ-\xI) + \sl*(\xI-\xJ) - 0.06+\sr *2}) ;
    
\draw[back]  ({\xJ}, {\srr*(\xK-\xI) + \sll*(\xJ-\xK) - 0.08}) 
  --  ({\xJ-2}, {\srr*(\xK-\xI) + \sll*(\xJ-\xK) - 0.08+2*\sr});

\draw[back]   ({\xK}, {\srr*(\xK-\xI) + \sll*(\xJ-\xK) - 0.08 + \srr*(\xK-\xJ)}) 
  --   ({\xK-2}, {\srr*(\xK-\xI) + \sll*(\xJ-\xK) - 0.08 + \srr*(\xK-\xJ)+2*\sr})  ;

\node[lbl, anchor=north east] at (OUT) {Transmitted};
\end{tikzpicture}

\vspace*{5mm}
        
\hspace*{-20mm}\begin{tikzpicture}[scale=0.85,>=stealth, on grid, auto]
\tikzset{
  state/.style={draw, rounded corners, minimum width=20mm, minimum height=7mm, font=\small},
  arrowstyle/.style={->, line width=1.2pt},
  layerbox/.style={draw, rounded corners, fill=gray!10, inner sep=8pt},
  openboxnode/.style={rounded corners, inner sep=10pt},  openedge/.style={draw, thick, dashed},                  initstate/.style={state, very thick, draw=black, fill=yellow!15} }

\node[initstate] (a0)    at (-0.4,   0.0) {$a_{0}$};          \node[state]     (d1)    at (3.2,   0.0) {$d_{1}$};
\node[state]     (a1)    at (6.4,   0.0) {$a_{1}$};
\node[state]     (d2)    at (9.6,   0.0) {$d_{2}$};
\node[state]     (a2)    at (12.8,  0.0) {$a_{2}$};
\node[state]     (d3a3)  at (16.4,  0.0) {$d_{3}=a_{3}$};     

\node[state]     (c0b0)  at (-0.4,  -3.0) {$c_{0}=b_{0}$};     \node[state]     (b1)    at (3.2,  -3.0) {$b_{1}$};
\node[state]     (c1)    at (6.4,  -3.0) {$c_{1}$};
\node[state]     (b2)    at (9.6,  -3.0) {$b_{2}$};
\node[state]     (c2)    at (12.8, -3.0) {$c_{2}$};
\node[initstate] (b3)    at (16.4, -3.0) {$b_{3}$};           

\begin{scope}[on background layer]
\node[layerbox, fit=(d1)(a1)(b1)(c1)] (L1box) {};
\node[layerbox, fit=(d2)(a2)(b2)(c2)] (L2box) {};
\node[openboxnode, draw=none, fit=(a0)(c0b0)] (M3box) {};
\node[openboxnode, draw=none, fit=(d3a3)(b3)] (M0box) {};
\end{scope}

\node[font=\footnotesize, anchor=south] at (L1box.north) {Layer 1: $n_1$, $h_1$};
\node[font=\footnotesize, anchor=south] at (L2box.north) {Layer 2: $n_2$, $h_2$};

\node[font=\footnotesize, anchor=south] at (M0box.north) {Medium 3: $n_3$};
\node[font=\footnotesize, anchor=south] at (M3box.north) {Medium 0: $n_0$};

\node[font=\scriptsize,  anchor=south] at (a0.north) {Init};
\node[font=\scriptsize,  anchor=north] at (b3.south) {Init};

\draw[openedge] (M0box.north west) -- (M0box.north east);
\draw[openedge] (M0box.south west) -- (M0box.south east);
\draw[openedge] (M0box.south west) -- (M0box.north west);

\draw[openedge] (M3box.north west) -- (M3box.north east);
\draw[openedge] (M3box.north east) -- (M3box.south east);
\draw[openedge] (M3box.south west) -- (M3box.south east);

\path[arrowstyle] (a0)   edge node {$t_{01}$} (d1);
\path[arrowstyle] (d1)   edge node {$p_{1}$}  (a1);
\path[arrowstyle] (a1)   edge node {$t_{12}$} (d2);
\path[arrowstyle] (d2)   edge node {$p_{2}$}  (a2);
\path[arrowstyle] (a2)   edge node {$t_{23}$} (d3a3);

\path[arrowstyle] (c1)   edge node {$p_{1}$}  (b1);
\path[arrowstyle] (b1)   edge node {$t_{10}$} (c0b0);

\path[arrowstyle] (c2)   edge node {$p_{2}$}  (b2);
\path[arrowstyle] (b2)   edge node {$t_{21}$} (c1);

\path[arrowstyle] (b3)   edge node {$t_{32}$} (c2);

\path[arrowstyle] (a0)  edge node [right] {$r_{01}$} (c0b0);
\path[arrowstyle] (b1)  edge node [right] {$r_{10}$} (d1);

\path[arrowstyle] (a1)  edge node [right] {$r_{12}$} (c1);
\path[arrowstyle] (b2)  edge node [right] {$r_{21}$} (d2);

\path[arrowstyle] (a2)  edge node [right] {$r_{23}$} (c2);
\path[arrowstyle] (b3)  edge node [right] {$r_{32}$} (d3a3);
\end{tikzpicture}
     \end{center}
\caption{Two layer structure $(0|1|2|3)$. In the top figure the arrows indicating partial waves and their coefficients. The bottom figure shows the multiple scattering and propagation events as a state machine.
    }
    \label{fig:layer}
\end{figure}

However, when three interfaces are present $(0|1|2|3)$, there exist particular instances where the Airy formula results in divergence when calculating the overall reflection and transmission coefficients (see Figs. \ref{fig:1}c,d). Three possible orderings of the partial wave terms in the Airy formula are:
\begin{align}
T_{023} & =T_{02} \, p_{2} \, t_{23}  \sum_{n=0}^{\infty} \big( p_2^2 r_{23}R_{20} \big)^n \label{T023} \\
T_{013} & = t_{01} \, p_{1} \, T_{13}  \sum_{n=0}^{\infty} \big( p_1^2 r_{10}R_{13}\big)^n \label{T013}\\
T_{03} &= \underbrace{t_{01} \, p_{1} \, t_{12} \, p_{2} \, t_{23}}_{\text{direct path}} \Bigg[
1 + \sum_{n=1}^{\infty}\;\sum_{\mathbf{b} \in \{1,2\}^n}
\Bigg(
\prod_{i=1}^n
\underbrace{p_{b_i}^2 \, r_{b_i b_{i-1}} \, r_{b_i b_{i+1}}}_{\text{round trip in layer } b_i}
\cdot
\prod_{i=1}^{n-1}
\underbrace{t_{b_i b_{i+1}}}_{\text{switch layer}}
\Bigg)
\Bigg]\label{T03}
\end{align}
with $b_0=0$ and $b_{n+1}=3$ and $t_{ii}=1$. These three representations correspond, respectively, to a nested summation where all scattering interactions are taken into account in the first layer (\ref{T023}), in the second layer (\ref{T013}) or the partial waves are added through a multipath summation (\ref{T03}) by number of interactions (see Fig. \ref{fig:layer}). In the presence of evanescence, these three formulas do not all converge.

\subsection{Optical Example of Superposition Breakdown}

We examine a scenario in optics where divergence in wave superposition arises. The system consists of scalar waves, representing linearly polarised light (p-polarisation) propagating through a plane-parallel two-layer structure with three interfaces. The angle of incidence is chosen near the critical angle for both layers, with one layer supporting propagating modes and the other supporting evanescent modes. This configuration is known to produce strong multiple-scattering effects and challenges conventional modal decomposition.

To generalise the Airy formula, we describe each partial wave in the multiple-scattering series as dependent on the amplitudes from previous scattering events. The partial wave amplitudes in layer \( i \) after $n$ scattering events are expressed using the interface scattering matrix and a propagation matrix:
\begin{equation}
\begin{bmatrix}
b_i^{n+1} \\
a_i^{n+1}
\end{bmatrix}
=
\begin{bmatrix}
0 & p_i \\
p_i & 0
\end{bmatrix}
\begin{bmatrix}
d_{i}^{n+1} \\
c_{i}^{n+1}
\end{bmatrix},
\quad\quad\quad
\begin{bmatrix}
c_i^{n+1} \\
d_{i+1}^{n+1}
\end{bmatrix}
=
\begin{bmatrix}
r_{i,i+1} & t_{i+1,i} \\
t_{i,i+1} & r_{i+1,i}
\end{bmatrix}
\begin{bmatrix}
a_{i}^n \\
b_{i+1}^n
\end{bmatrix}\label{rec}
\end{equation}
where \( a_i \) and \( b_i \) are the amplitudes of the partial waves in layer $i$ in the forward and backward direction, with the electric field $E^n_y$ and magnetic field $H^n_x$ in layer $i$ defined by:
\begin{align}
E^n_y(z_i) &= e^{\imath\, (k_x x-\omega t)} \big(a^n_i e^{\imath\, k_{z,i} z_i}+b^n_i e^{-\imath\, k_{z,i} z_i}\big) \\
H^n_x(z_i) &= -\frac{k_{z,i}}{\omega \mu_0}e^{\imath\, (k_x x-\omega t)} \big(a^n_i e^{\imath\, k_{z,i} z_i}-b^n_i e^{-\imath\, k_{z,i} z_i}\big) 
\end{align}
where $\mu_0$ is the permeability of the free space.

The field is the sum between the two waves with a normal wavevector component $k_{z,i}=\sqrt{n_i^2\omega^2/c^2-k_x^2}$ where we use a local coordinate system $z_i$ with its origin at the left interface. When $k_{z,i}$ is real, there are two travelling waves; if $k_{z,i}$ is imaginary, then the two waves are evanescent. The optical frequency $\omega$ and the transverse wavevector $k_x$ are defined by the incident wave. In this more generic multi-scattering description, the Airy formula is:
\begin{equation}
a_i = \sum_{n=1}^\infty a^n_i, \quad b_i =\sum_{n=1}^\infty b^n_i
\label{gen-airy}
\end{equation}
corresponding to the total wave amplitudes after all multiple-scattering events have been taken into account. 

A one-step scattering event for the whole structure can be represented as a single-matrix multiplication that acts on all partial-wave amplitudes in all layers at each iteration. The convergence of the series (\ref{gen-airy}) and of the multipath summation (\ref{T03}) is guaranteed only when all the eigenvalues of the resulting event scattering matrix have magnitudes smaller than one. An eigenvalue equal to one corresponds to a guided mode, while eigenvalues greater than one lead to divergence in the Airy-type series. 
Spectral eigenvalue conditions on the event (or interaction) operator are the layered analogue of the Foldy-Lax criteria for the coherent field and effective parameters in random media,
where breakdowns of naive series occur beyond leading density or near resonant configurations
\cite{foldy1945multiple}.

In the specific case of a three-interface structure $(n_2|n_1|n_2|n_1)$ composed of two mediums with refractive indices \( n_1 \) and \( n_2 \), we analytically determine the four eigenvalues of the event scattering matrix.
\begin{align}
    \lambda_{1,...,4} & = \pm\left(\frac{1}{2}(p_1^2+p_2^2)r_{12}^2\pm\frac{1}{2}\sqrt{\left(p_1^2+p_2^2\right)^2r_{12}^4-4 p_1^2p_2^2r_{12}^2}\right)^{1/2} \label{eq:lambda4}
\end{align}
These eigenvalues are determined by assembling the $4\times4$ block event matrix ${\bf E}={\bf P\,S}$ acting on $(a_1,b_1,a_2,b_2)$. The roots of the characteristic polynomial of $E$ are defined by \eqref{eq:lambda4} (for details, see the Methods section).

\begin{figure}
    \centering
    \includegraphics[width=0.95\linewidth]{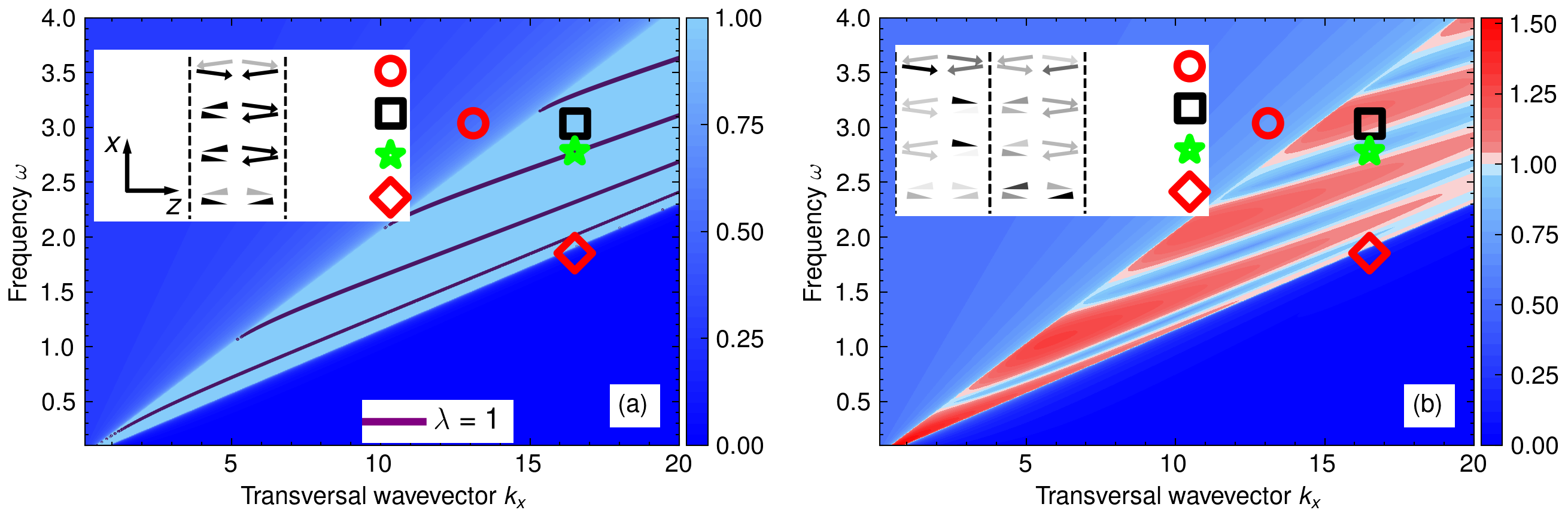}
    \caption{Maximum eigenvalue magnitude \(|\lambda|\) of the event scattering matrix for  (a) single- and (b) two-layer structures. The insets show the eigenmodes for the four cases considered. The arrows represent propagating waves while the triangles evanescent waves. Where red zones correspond to divergent domains. The shade of the arrows and triangles are signifying the amplitudes.}
    \label{fig:2}
\end{figure}

Figure~\ref{fig:2} shows the maximum magnitude of the eigenvalue \(|\lambda|\) for the single- and two-layer structures considered and defined in Figure~\ref{fig:1}. The analytic expression for the two-layer structure demonstrates the existence of divergence regions for specific wavelength and transverse wavevector values in the three-layer structure. In contrast, as anticipated, there is no divergence for the single-layer case. In fact, for the single-layer structure, we have $\lambda_{1,2}=\pm \imath \, p_1 r_{12}$.

\section{Resolving divergence using power flux eigenmodes}

The divergence observed in multilayered systems originates from the inability to properly normalise evanescent and inhomogeneous wave components. Unlike propagating modes, which carry finite energy and can be normalised with respect to conserved quantities such as power flux, evanescent waves decay exponentially and do not transport energy across layers.
By enforcing orthogonality with respect to the normal component of the Poynting vector,
our basis inherits the energy-conservation and unitarity structure familiar from guided‑wave and resonator theory
\cite{Marcuse1991}.
This makes them non-normalisable in the conventional sense, especially when they appear in infinite series expansions like those used in multiple scattering formulations.

In such cases, the standard modal decomposition fails to converge because the basis functions do not form a complete orthonormal set with respect to the relevant conserved quantity, typically energy or power flux. This issue is exacerbated when the system includes interfaces near or beyond the critical angle, where evanescent contributions dominate and interfere destructively or constructively in ways that amplify divergence.

To resolve this divergence problem, we must reconsider the basis used to represent the wave field. Instead of relying on conventional plane wave or field amplitude modes, we introduce eigenmodes of the power flux measure. These modes are constructed to be orthonormal with respect to the power flux, ensuring that each contributes a finite, well-defined amount of energy without interference effects. As a consequence of this mode definition, the overall superposition remains physically meaningful and mathematically convergent. This approach not only restores convergence, but also aligns with the physical intuition that only modes contributing to measurable quantities (like transmitted or reflected power) should be retained in the representation. 

The power flux eigenmode basis is defined by the normal component of the Poynting vector 
\begin{equation}
    P_z=-\frac{1}{2}\big(H^*_xE_y+H_xE^*_y\big) =
    \begin{bmatrix}
E_y^* ,
H_x^*
\end{bmatrix}
    \begin{bmatrix}
0 & -\frac{1}{2} \\
-\frac{1}{2} & 0
\end{bmatrix}
\begin{bmatrix}
E_y \\
H_x
\end{bmatrix}
\end{equation}
which has two eigenmodes defined by the y-component of the electric fields:
\begin{align}
E^{\pm}_y(z) &=\frac{\sqrt{\omega\mu_0}e^{\imath\, (k_x x-\omega t)}}{2|k_z|}\big( (\sqrt{k_z}\pm\sqrt{k^*_z} ) e^{\imath\, k_z z}+(\sqrt{k_z}\mp\sqrt{k^*_z} ) e^{-\imath\, k_z z}\big) \\
    &=\sqrt{\omega\mu_0}e^{\imath\, (k_x x-\omega t)}\left( \frac{\cos(k_z z)}{\sqrt{k^*_z}} \pm \imath\,\frac{\sin(k_z z)}{\sqrt{k_z}}\right)
\end{align}
where we have normalised the eigenmodes with respect to the normal power flux at $z=0$. Note that these solutions of the wave equation diverge at infinity for evanescent and inhomogeneous normal waves. 

Using these power eigenmodes we define the field in the $i$-th layer as a superposition: 
\begin{equation}
E^n_y(z_i) = \tilde a^n_i E^+_y(z_i)+\tilde b^n_i E^-_y(z_i)
\end{equation}
where $\tilde a_i^{n+1}$ and $\tilde b_i^{n+1}$ are the amplitudes of the two modes in the layer $i$ after scattering events $n$. If the $i$-th layer is semi-infinite, then non-reflecting natural boundary conditions, equivalent to a decaying field correspond to $-(\sqrt{k_z}\pm\sqrt{k^*_z} )\tilde a^n_i=(\sqrt{k_z}\mp\sqrt{k^*_z} )\tilde b^n_i$ where the signs depend on the direction of the semi-infinite space.

Using these power eigenmodes we can define the scattering and propagation matrices as: 
\begin{equation}
\begin{bmatrix}
\tilde b_i^{n+1} \\
\tilde a_i^{n+1}
\end{bmatrix}
=
\begin{bmatrix}
\tilde q_i & \tilde p_i \\
\tilde p_i & \tilde q_i
\end{bmatrix}
\begin{bmatrix}
\tilde d_{i}^{n+1} \\
\tilde c_{i}^{n+1}
\end{bmatrix},
\quad\quad\quad
\begin{bmatrix}
\tilde c_i^{n+1} \\
\tilde d_{i+1}^{n+1}
\end{bmatrix}
=
\begin{bmatrix}
\tilde r_{i,i+1} & \tilde t_{i+1,i} \\
\tilde t_{i,i+1} & \tilde r_{i+1,i}
\end{bmatrix} 
\begin{bmatrix}
\tilde a_{i}^n \\
\tilde b_{i+1}^n
\end{bmatrix}\label{rec-tilde}
\end{equation}
with
\begin{align*}
    \tilde t_{ij} &= \frac{2\sqrt{k_{z,i}k_{z,j}^*}}{|k_{z,i}|+|k_{z,j}|} &   
     \tilde q_{i} &= \frac{\imath\,(k_{z,i}-k_{z,i}^*) \sin(h_i k_{z,i})}{2|k_{z,i}|\cos(h_i k_{z,i})-\imath\, (k_{z,i}+k_{z,i}^*) \sin(h_i k_{z,i})} \\
     \tilde r_{ij} &= \frac{|k_{z,i}|-|k_{z,j}|}{|k_{z,i}|+|k_{z,j}|} &
       \tilde p_{i} &= \frac{2|k_{z,i}|}{2|k_{z,i}|\cos(h_i k_{z,i})-\imath\, (k_{z,i}+k_{z,i}^*) \sin(h_i k_{z,i})}
\end{align*}
Before using these equations, we note that for real $k_z$ the recursive event matrices described by (\ref{rec-tilde}) are identical to (\ref{rec}). This is not the case for complex-valued $k_z$. Another important property is that the interface scattering matrix is unitary regardless of the value of $k_z$, while the magnitude of all eigenvalues of the propagation matrix are smaller than one for inhomogeneous waves and equal to one for propagating and evanescent waves (lossless and passive materials). It is these properties that lead to the convergence of Airy series. 

\begin{figure}
    \centering
    \includegraphics[width=0.95\linewidth]{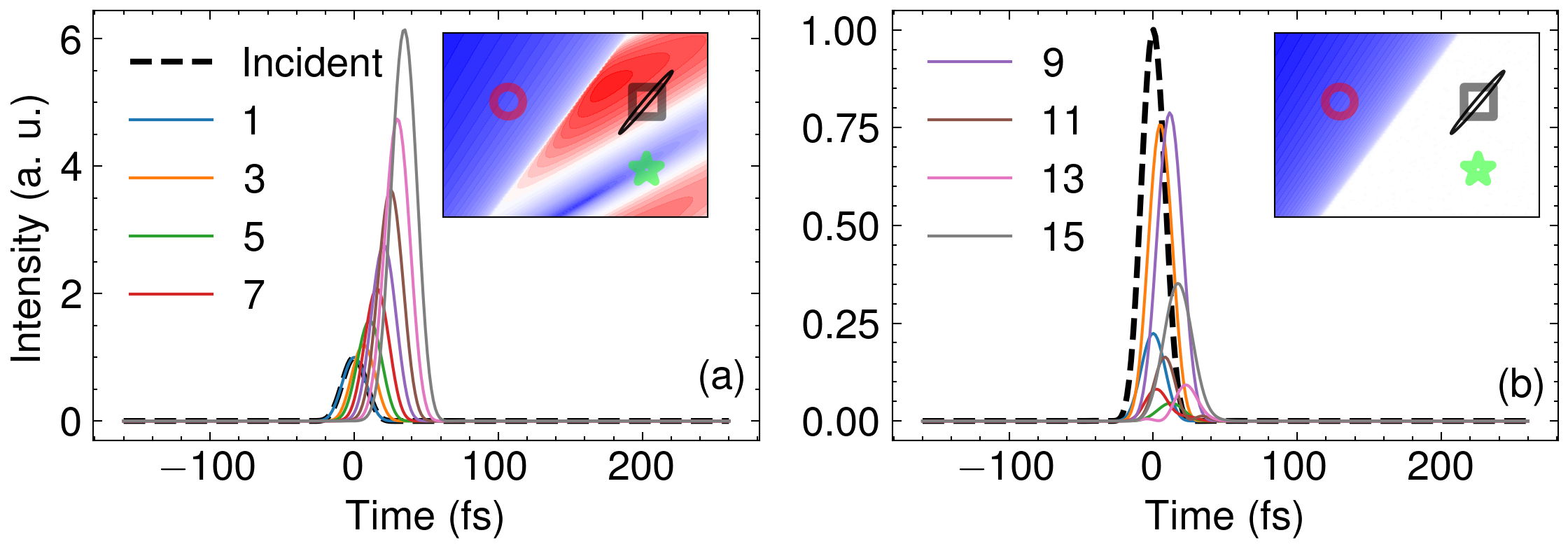}
    \caption{
Pulse reflected from the double layer structure at different event number using (a) the standard propagation mode definition and (b) the power mode definition. The finite pulse is defined in the evanescent case and its transversal wavevector and frequency spectrum is shown in the inset as a contour plot. }
    \label{fig:4}
\end{figure}

To illustrate numerically the effectiveness of the power-flux eigenmode approach, we used it to calculate the reflected field of an optical pulse. 
In this simulation, a spatially and temporally localised pulse is incident on the three-interface multilayer structure. Using the recursive scattering approach, the reflected field is reconstructed in the time domain after an odd number of events. Figure \ref{fig:4}a shows the divergence when using standard plane wave modes. This divergence disappears when the power flux eigenmodes are used, as shown in Figure \ref{fig:4}b.

\section{Methods}

\subsection{Numerical evaluation}
A numerical model based on the scattering matrix formalism was used to simulate the propagation of the monochromatic waves through multilayer dielectric structures. Each interface and medium are represented by a scattering matrix $2 \times 2$, corresponding to the Fresnel coefficients and the accumulation of the optical path phases, respectively. The overall scattering for a multilayer stack is derived through recursive matrix multiplication, where each recursion corresponds to one event step.

To be specific, we chose common optical dielectric materials for our examples. On the incident side of the system we consider a rutile film coupling prism (refraction index $n_2=2.59$ at 620~nm) followed by two coating layers on a glass substrate ($n_1=1.46$). 

To further clarify the scattering in multilayer structures, the maximal eigenvalue of the total scattering matrix was computed for the two configurations considered. For each combination of the transverse wavevector $k_x$ and frequency $\omega$, individual scattering matrices were combined to determine the total scattering matrix of the system. Its eigenvalues were calculated and the largest eigenvalue (in magnitude) was extracted to be used in contour plots that show the regions of convergence and divergence for the two structures considered (Figure \ref{fig:2}). 

To examine the temporal evolution, the incident pulse is modelled using an angular spectrum approach. This method allows for the construction of spatially and temporally localised wave packets by integrating over a range of incident angles corresponding to a range of transversal wavevector components. 
To study the divergence of the optical field within the multilayer systems considered, we determined the time-resolved intensity of the reflected field at a fixed spatial location. The reflected field was calculated by transformed back to the real space via inverse Fourier transforms of the individual plane waves reflected by the structure after a fixed number of events. Figure \ref{fig:4} shows the field intensity in the centre of the pulse ($x=0$) with an incident pulse at an angle of 38.9$^\circ$ corresponding to a propagating solution in medium $n_2=2.59$ (TiO$_2$) and an evanescent solution in medium $n_1=1.46$ (SiO$_2$) at a wavelength $\lambda=620$ nm. The thickness of the $n_1$ layer is $0.16\lambda$ and the thickness of the $n_2$ layer is $0.68\lambda$. We considered an ultra-short pulse of 14.5 periods with a waist of 60.8$\lambda$.

\subsection{Normalised power flux modes}
To ensure convergence in multilayer systems when the solution involves evanescent and inhomogeneous components, we introduce a basis of normalised power flux modes. These modes are defined as eigenfunctions of the power flux operator, specifically constructed to be orthonormal with respect to the normal component of the Poynting vector. Unlike conventional plane wave modes, which may fail to conserve energy in the presence of non-propagating waves, power flux modes guarantee that each basis function contributes a finite and measurable amount of energy. The electric field in each layer is expressed as a superposition of two such modes, corresponding to forward and backward energy flows. The associated scattering and propagation matrices are reformulated in this basis, preserving unitarity and ensuring that all eigenvalues remain bounded. This framework allows for stable recursive computation of wave amplitudes across multiple scattering events and restores convergence.

\subsection{Interface and media scattering for power flux modes}
In the power-flux mode formalism, both interface and media scattering processes are reformulated to preserve energy conservation and ensure convergence in multilayer systems. Each interface is represented by a unitary scattering matrix constructed in the power flux eigenmode basis, where the reflection and transmission coefficients are derived from the z-components of the complex wavevectors. These coefficients differ from conventional Fresnel expressions by incorporating the power-normalised eigenfunctions of the Poynting vector. Similarly, propagation through each medium is described by a matrix whose eigenvalues are bounded, equal to one for propagating and evanescent waves and less than one for inhomogeneous waves, ensuring stability across recursive scattering iterations. The field in each layer is expressed as a superposition of two power-flux modes, and their amplitudes are updated through matrix operations that couple interface and propagation effects. This approach guarantees that all wave components contribute finite energy, resolving divergence issues, and enabling accurate modelling of complex multilayer structures.

\subsection{Global event scattering matrix}

To address the divergence issues inherent in conventional modal decompositions of multilayered wave systems, we constructed a global scattering matrix using power flux eigenmodes. These modes are orthonormal with respect to the conserved quantity of normal power flux, ensuring that each contributes a finite and physically meaningful amount of energy. The scattering matrices for individual interfaces and propagation through media were reformulated in this basis, preserving unitarity and guaranteeing that all eigenvalues of the propagation matrices remain bounded. Recursive matrix multiplication was employed to simulate multiple scattering events, with each iteration updating the amplitudes of the power flux modes. This approach ensures convergence of the series expansion even in the presence of evanescent and inhomogeneous waves. The global event scattering matrix, formed by combining all interface and propagation matrices, was analysed through its eigenvalue spectrum to identify regions of convergence and divergence. This formalism was applied to simulate the reflection of ultra-short pulses in multilayer structures, demonstrating stable and physically interpretable results in contrast to the divergent behaviour observed with standard plane wave modes.

The interface scattering matrices can be collected and written as a block matrix: \[
\begin{bmatrix}
\tilde{d}_1^{n+1} \\
\tilde{c}_1^{n+1} \\
\tilde{d}_2^{n+1} \\
\tilde{c}_2^{n+1} \\
\tilde{d}_3^{n+1} \\
\vdots \\
\tilde{c}_n^{n+1} \\
\tilde{d}_{n+1}^{n+1} \\
\tilde{c}_{n+1}^{n+1}
\end{bmatrix}
=
\begin{bmatrix}
\tilde{r}_{1} & 0 & 0 & 0 & 0 & \cdots & 0 & 0 & 0 \\
0 & \tilde{r}_{1,2} & \tilde{t}_{2,1} & 0 & 0 & \cdots & 0 & 0 & 0 \\
0 & \tilde{t}_{1,2} & \tilde{r}_{2,1} & 0 & 0 & \cdots & 0 & 0 & 0 \\
0 & 0 & 0 & \tilde{r}_{2,3} & \tilde{t}_{3,2} & \cdots & 0 & 0 & 0 \\
0 & 0 & 0 & \tilde{t}_{2,3} & \tilde{r}_{3,2} & \cdots & 0 & 0 & 0 \\
\vdots & \vdots & \vdots & \vdots & \vdots & \ddots & \vdots & \vdots & \vdots \\
0 & 0 & 0 & 0 & 0 & \cdots & \tilde{r}_{n,n+1} & \tilde{t}_{n+1,n} & 0 \\
0 & 0 & 0 & 0 & 0 & \cdots & \tilde{t}_{n,n+1} & \tilde{r}_{n+1,n} & 0 \\
0 & 0 & 0 & 0 & 0 & \cdots & 0 & 0 & \tilde{r}_{n+1} 
\end{bmatrix}
\begin{bmatrix}
\tilde{b}_1^n \\
\tilde{a}_1^n \\
\tilde{b}_2^n \\
\tilde{a}_2^n \\
\tilde{b}_3^n \\
\vdots \\
\tilde{a}_n^n \\
\tilde{b}_{n+1}^n \\
\tilde{a}_{n+1}^n 
\end{bmatrix}
=
{\bf S}
\begin{bmatrix}
\tilde{b}_1^n \\
\tilde{a}_1^n \\
\tilde{b}_2^n \\
\tilde{a}_2^n \\
\tilde{b}_3^n \\
\vdots \\
\tilde{a}_n^n \\
\tilde{b}_{n+1}^n \\
\tilde{a}_{n+1}^n 
\end{bmatrix}
\]
where $\tilde r_1$ and $\tilde r_{n+1}$ denote the boundary reflectivity on the left and right sides of the multilayered structure, respectively.
The subsequent step of the event is characterised by the use of propagation matrices that connect the scattered fields to the incident fields at the interface through:

\[
\begin{bmatrix}
\tilde b_1^{n+1} \\
\tilde a_1^{n+1} \\
\tilde b_2^{n+1} \\
\tilde a_2^{n+1} \\
\vdots \\
\tilde b_{n+1}^{n+1} \\
\tilde a_{n+1}^{n+1}
\end{bmatrix}
=
\begin{bmatrix}
\tilde q_1 & \tilde p_1 & 0 & 0 & \cdots & 0 & 0 \\
\tilde p_1 & \tilde q_1 & 0 & 0 & \cdots & 0 & 0 \\
0 & 0 & \tilde q_2 & \tilde p_2 & \cdots & 0 & 0 \\
0 & 0 & \tilde p_2 & \tilde q_2 & \cdots & 0 & 0 \\
\vdots & \vdots & \vdots & \vdots & \ddots & \vdots & \vdots \\
0 & 0 & 0 & 0 & \cdots & \tilde q_{n+1} & \tilde p_{n+1} \\
0 & 0 & 0 & 0 & \cdots & \tilde p_{n+1} & \tilde q_{n+1}
\end{bmatrix}
\begin{bmatrix}
\tilde d_1^{n+1} \\
\tilde c_1^{n+1} \\
\tilde d_2^{n+1} \\
\tilde c_2^{n+1} \\
\vdots \\
\tilde d_{n+1}^{n+1} \\
\tilde c_{n+1}^{n+1}
\end{bmatrix}
=
{\bf P}
\begin{bmatrix}
\tilde d_1^{n+1} \\
\tilde c_1^{n+1} \\
\tilde d_2^{n+1} \\
\tilde c_2^{n+1} \\
\vdots \\
\tilde d_{n+1}^{n+1} \\
\tilde c_{n+1}^{n+1}
\end{bmatrix}
\]

The global event scattering matrix is the result of the matrix multiplication described by ${\bf PS}$, which determines the change in amplitudes, $\tilde a_i$ and $\tilde b_i$, for forward and backward modes during a recursive scattering event.

In the special case of $(n_2|n_1|n_2|n_1)$ we have:
\[
{\bf S} = \begin{bmatrix}
-\tilde r_{12} & 0 & 0 & 0 \\
0 & -\tilde r_{12} & \tilde t_{12} & 0 \\
0 & \tilde t_{21} & \tilde r_{12} & 0 \\
0 & 0 & 0 & \tilde r_{12}
\end{bmatrix}
\]

\[
{\bf P} = \begin{bmatrix}
\tilde q_2 & \tilde p_2 & 0 & 0 \\
\tilde p_2 & \tilde q_2 & 0 & 0 \\
0 & 0 & \tilde q_1 & \tilde p_1 \\
0 & 0 & \tilde p_1 & \tilde q_1
\end{bmatrix}
\]

\[
{\bf P} \cdot {\bf S} = 
\begin{bmatrix}
-\tilde q_2 \tilde r_{12} & -\tilde p_2 \tilde r_{12} & \tilde p_2 \tilde t_{12} & 0 \\
-\tilde p_2 \tilde r_{12} & -\tilde q_2 \tilde r_{12} & \tilde q_2 \tilde t_{12} & 0 \\
0 & \tilde q_1 \tilde t_{21} & \tilde q_1 \tilde r_{12} & \tilde p_1 \tilde r_{12} \\
0 & \tilde p_1 \tilde t_{21} & \tilde p_1 \tilde r_{12} & \tilde q_1 \tilde r_{12}
\end{bmatrix}
\]
which for $\tilde q_1=\tilde  q_2=0$ has an analytic solution for its eigenvalues \eqref{eq:lambda4}. 

\subsection{Numerical spectral decomposition of ultrashort Gaussian pulses}
To investigate the temporal dynamics and convergence behaviour of wave propagation in multilayered media, we performed a numerical spectral decomposition of ultrashort Gaussian pulses using an angular spectrum approach. This method constructs spatially and temporally localised wave packets by integrating over a range of incident angles, corresponding to a distribution of transverse wavevector components. Each spectral component was propagated through the multilayer structure using the global scattering matrix formalism, and the reflected field was reconstructed via inverse Fourier transforms. The decomposition allowed us to resolve the time-dependent intensity of the reflected pulse at a fixed spatial location, revealing divergence when standard plane wave modes were used. In contrast, when power flux eigenmodes were employed, the decomposition yielded stable and convergent results, demonstrating the physical validity of the power-mode basis in capturing energy-conserving dynamics even in the presence of evanescent and inhomogeneous wave components.

\section*{Discussion and Conclusion}

This study demonstrates that not all basis sets are physically equivalent in linear wave systems, even though the governing equations remain mathematically linear. 
A closely related sensitivity to representation arises in periodic media, where Bloch-mode completeness, interface matching, and evanescent content govern device‑level behaviour
\cite{joannopoulos2008photonic}.
Conventional bases, such as those constructed from arbitrary field components like plane waves or modal amplitudes, often fail to preserve essential physical properties such as energy conservation when evanescent or inhomogeneous waves are present. This failure becomes particularly evident in multilayered systems, where the superposition principle appears to break down due to divergence in infinite series expansions. Importantly, this divergence is not a numerical artefact but a fundamental consequence of using non-normalisable components in the representation.

The breakdown of the superposition principle in these systems highlights a critical limitation of traditional modal decomposition techniques. 
In strongly disordered settings, repeated multiple scattering can even suppress transport altogether (Anderson localisation),
underscoring that convergence and normalisation issues in wave superpositions are not merely technicalities, but can signal qualitative transport changes
\cite{anderson1958absence}.
Evanescent waves, which decay exponentially and do not transport energy across layers, cannot be normalised in the same way as propagating modes. When these components are included in infinite series expansions, the lack of a proper normalisation framework leads to divergence. This observation underscores the need for physically motivated basis sets that respect conserved quantities such as energy or power flux.

To address this issue, we introduced power-flux eigenmodes as an alternative representation. These modes form an orthonormal basis with respect to the power flux, ensuring that each mode contributes a finite and physically meaningful amount of energy. By construction, this approach guarantees convergence in series expansions and restores the physical interpretability of the solutions. Furthermore, the unitary nature of the interface scattering matrices and the bounded eigenvalues of the propagation matrices in the power-flux basis provide mathematical stability, even in scenarios involving strong multiple scattering or near-critical angle incidence.

The implications of this framework extend beyond optics. The formalism generalises naturally to other wave domains, including acoustics, elastodynamics, and quantum mechanics. In quantum systems, for example, similar divergence issues arise in tunnelling and bound-state problems, where unitarity and probability conservation are essential. Moreover, the ability to incorporate evanescent and inhomogeneous components without sacrificing normalisability opens pathways for modelling phenomena such as quantum tunnelling, near-field coupling, and wave localisation using path integrals without the need for renormalisation or regularisation techniques.


\backmatter

\bmhead{Supplementary information}

The research data underpinning this publication can be accessed at \url{https://doi.org/10.17630/da4b8bd9-0af6-48f7-bb47-babd40ff038c} \cite{supmat}.  


\section*{Declarations}
Not applicable





\begin{appendices}

\section{Other methods to remove divergences}

When the multiple-scattering series for a lossless multilayer optical system diverges, assigning a finite value through analytic continuation is a purely mathematical device without physical causality. The continuation constructs a value by extending the analytic domain of the convergent series, but the actual system cannot realise this process dynamically as it does not sum the divergent series to reach that point. Similarly, enforcing convergence by reordering the terms of the series is an arbitrary mathematical manipulation that violates the causality of the system. Introducing a small absorption term and extrapolating to zero absorption is often considered more physical because it corresponds to realisable lossy systems where the series converges. However, if the true system is strictly non-absorbing, this procedure still relies on a family of hypothetical systems that do not exist in reality. In all of these cases, the assigned value emerges from the analytic structure of the model rather than from the causal evolution of the physical system. Thus, although such regularisations are often pragmatically justified because they yield correct measurable predictions, they lack physical interpretability, as they do not represent what the system actually does.

\subsection{Regularisation using Abel summation and analytic continuation}

\begin{figure}[h!]
\centering
\includegraphics[width=0.7\linewidth]{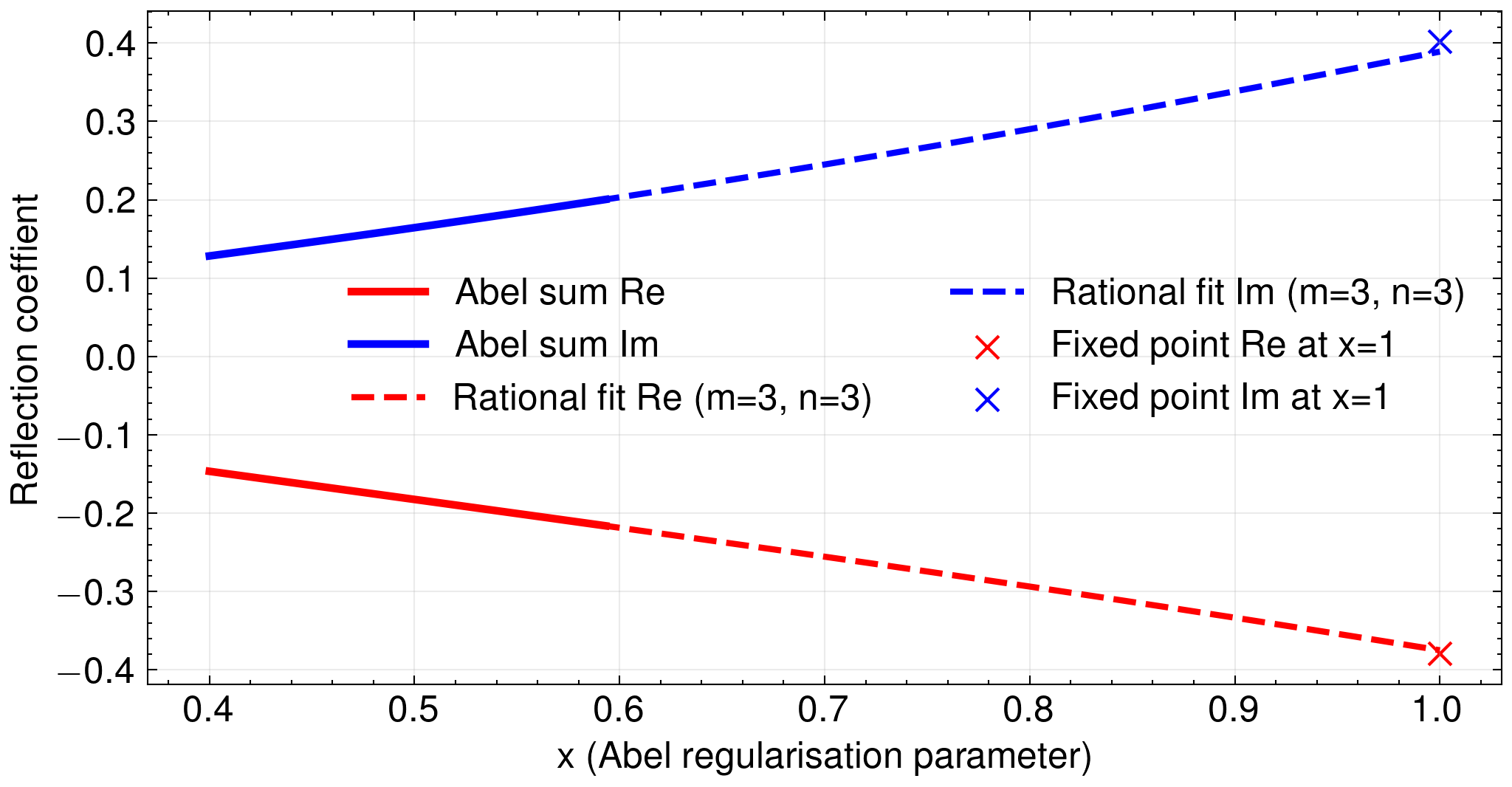}
\caption{Reflected field coefficient for the structure discussed in figure (Fig. \ref{fig:2}).
Solid curves: real (red) and imaginary (blue) parts of the  Abel sum $\mathbf{R}(x)$ for $x<x_{max}$.
Dashed curves: rational fit (Padé-like) providing analytic continuation toward $x=1$.
Crosses: fixed-point solution $(\mathbf{I}-\mathbf{P\cdot S})^{-1}\mathbf{s}_0$ at $x=1$.}
\label{fig:abel_fit}
\end{figure}

Consider the divergent Neumann series:
\begin{equation}
{\bf R}=\sum_{k=0}^{\infty} \big(\mathbf{P\cdot S}\big)^k \mathbf{s}_0,
\end{equation}
where $\mathbf{P\cdot S}\in\mathbb{C}^{N\times N}$ is the round-trip operator and $\mathbf{s}_0$ is the initial state.
As we discussed in the main body of the paper, when $\lambda_{max}=\max(\rho(\mathbf{P\cdot S})) \ge 1$, this series does not converge in the usual sense.
To regularise it, we introduce an Abel parameter $x\in[0,1)$ and define
\begin{equation}
\mathbf{R}(x) = \sum_{k=0}^{\infty} (x\; \mathbf{P\cdot S})^k \mathbf{s}_0,
\end{equation}
which converges for $x< \frac{1}{\lambda_{max}}=x_{max}$ and admits the closed form
\begin{equation}
\mathbf{R}(x) = (\mathbf{I}-x\;\mathbf{P\cdot S})^{-1}\mathbf{s}_0.
\end{equation}
The analytic continuation of the Abel sum corresponds to the regularised original Neumann series
\begin{equation}
\lim_{x\to 1^-} \mathbf{R}(x).
\end{equation}
Numerically, for each convergent $x\in[x_{\min},x_{\max}]$, we compute the Abel sum $\mathbf{R}(x) $. To approximate the limit $x\to 1^-$ and extend beyond the sampled region, we fit a rational function:
\begin{equation}
\hat{y}(x) = \frac{P_m(x)}{Q_n(x)}, \qquad
P_m(x)=\sum_{i=0}^{m}p_i x^i,\;\; Q_n(x)=1+\sum_{j=1}^{n}q_j x^j,
\end{equation}
to the coefficient of interest of the complex field from the state vector $\bf s$. 
Figure~\ref{fig:abel_fit} illustrates that the Neumann series diverges as $x\to 1$, while the Abel-summed form remains well defined for $x<x_{max}$.
The rational approximation accurately captures the trend and predicts the limiting value at $x=1$, even when the direct summation fails.

\subsection{Event Order Reversal}

\begin{figure}[!ht]
\centering
\includegraphics[width=0.9\linewidth]{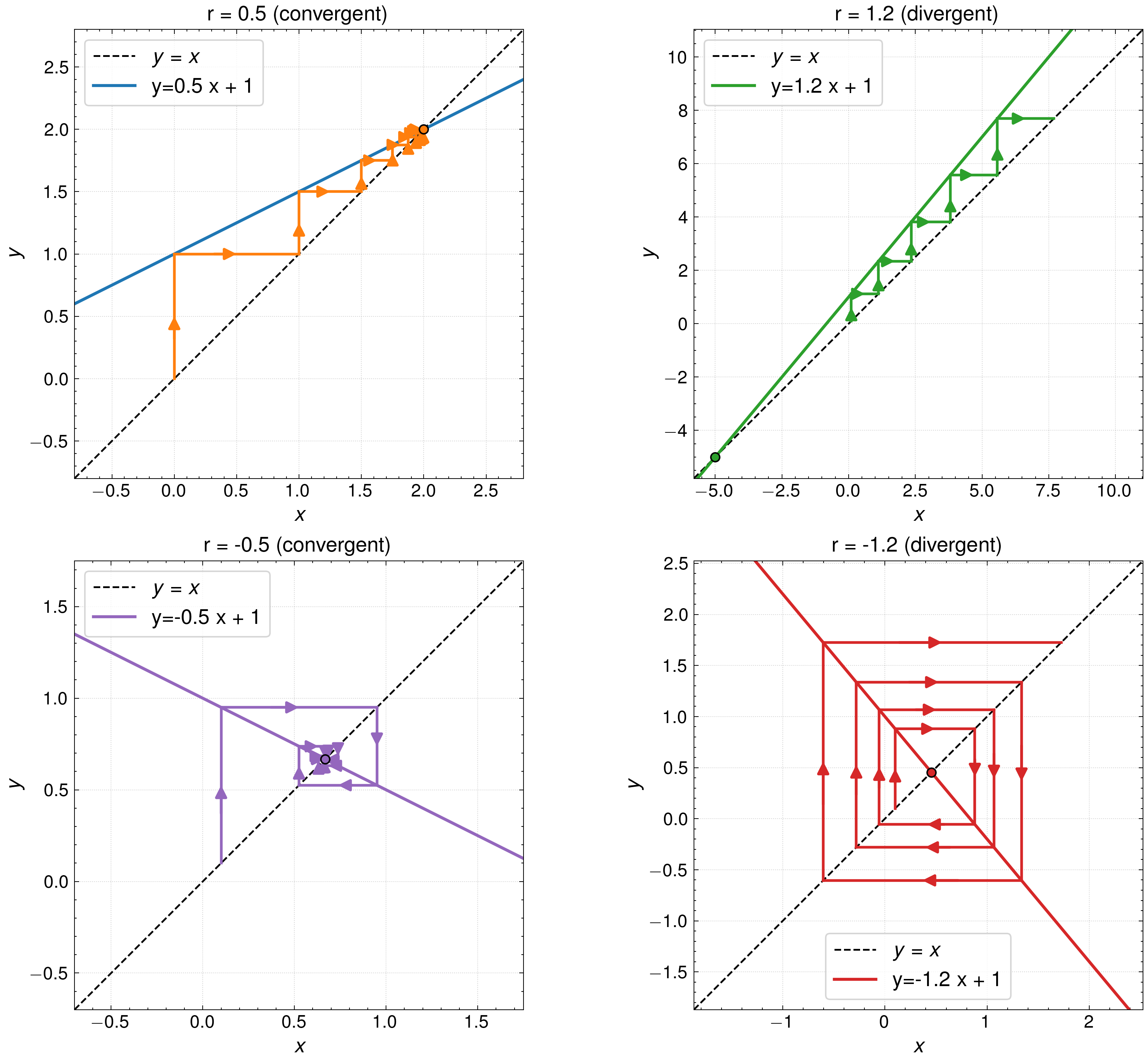}
\caption{
Each panel shows the iterative map $R_{k+1}=1+rR_k$ (or its reversed form) for different values of $r$.  
For $|r|<1$ (top row), the forward iteration converges to the fixed point $(1-r)^{-1}$, as indicated by arrows spiralling inward.  
For $|r|>1$ (bottom row), the forward iteration diverges (arrows move outward), but reversing the iteration changes the arrow direction and restores convergence to the fixed point.}
\label{fig:cobweb}
\end{figure}

For a scalar round-trip factor $r \in \mathbb{C}$, the multiple-scattering expansion reduces to the geometric Neumann series:
\[
R=\sum_{k=0}^{\infty} r^k = \frac{1}{1-r}, \qquad |r|<1.
\]
This series can be interpreted as the solution of the fixed-point equation:
\[
R = 1 + r R,
\]
which admits the unique solution $R = (1-r)^{-1}$ whenever $|r|<1$.  
An iterative scheme to approximate this fixed point is:
\[
R_{k+1} = 1 + r R_k,
\]
and its behaviour is illustrated in Figure~\ref{fig:cobweb}. For $|r|<1$, the arrows in the cobweb diagram show convergence toward the fixed point because the map is a contraction.  
When $|r|>1$, the same forward iteration diverges, and the arrows in the figure move away from the fixed point.  
However, reversing the iteration order yields the following result:
\[
R_{k+1} = r^{-1}(R_k - 1),
\]
which converges for $|r|>1$. In the cobweb diagrams, this reversal corresponds to reversing the direction of the arrows, indicating that the sequence now approaches the fixed point from outside the original convergence region.  
Physically, this reversal can be interpreted as inverting the causal order of scattering events, introducing a non-physical but mathematically consistent regularisation. 

For a series that involves scattering matrices such as $\bf P \cdot S$, the same principle applies after representing the round-trip matrix on its own eigenvector basis.   
In this case, the reversal is performed only for the eigenvectors whose eigenvalues satisfy $|\lambda|>1$, while the modes with $|\lambda|<1$ retain the forward iteration.  
This selective reversal ensures convergence toward the fixed point for all components.

\subsection{Physically based regularisation using absorption and analytic continuation}

In this approach, the divergent multiple-scattering series is regularised by introducing a small imaginary part into the refractive indices, corresponding to material absorption.  
Let $n_1, n_2 \in \mathbb{C}$ denote the refractive indices of the two layers. We replace
\begin{equation}
n_j \;\mapsto\; n_j - \imath\,\alpha, \qquad \alpha > 0,
\end{equation}
where $\alpha$ is an absorption parameter. For sufficiently large absorption parameters, this modification ensures that the round-trip operator $\mathbf{P(\alpha)\cdot S(\alpha)}$ has a spectral radius smaller than one such that the Neumann series converges. Physically, $\alpha$ represents a controlled loss that damps high-order scattering paths.
The original (lossless) system corresponds to $\alpha = 0$. We compute the convergent sum:
\begin{equation}
\mathbf{R}(\alpha) = \sum_{k=0}^{\infty} \big(\mathbf{P(\alpha)\cdot S(\alpha)}\big)^k \mathbf{s}_0 = \big(\mathbf{I}-\mathbf{P(\alpha)\cdot S(\alpha)}\big)^{-1}\mathbf{s}_0,
\end{equation}
for a sequence of decreasing $\alpha$ values. 
The desired value for the lossless system is then interpreted as
\begin{equation}
\lim_{\alpha \to 0^+} \mathbf{R}(\alpha),
\end{equation}
analogous to Abel summation but with a physically motivated regularisation parameter. This can be calculated using analytic continuation, which numerically can be estimated using rational function fit. 

\begin{figure}[h!]
\centering
\includegraphics[width=0.7\linewidth]{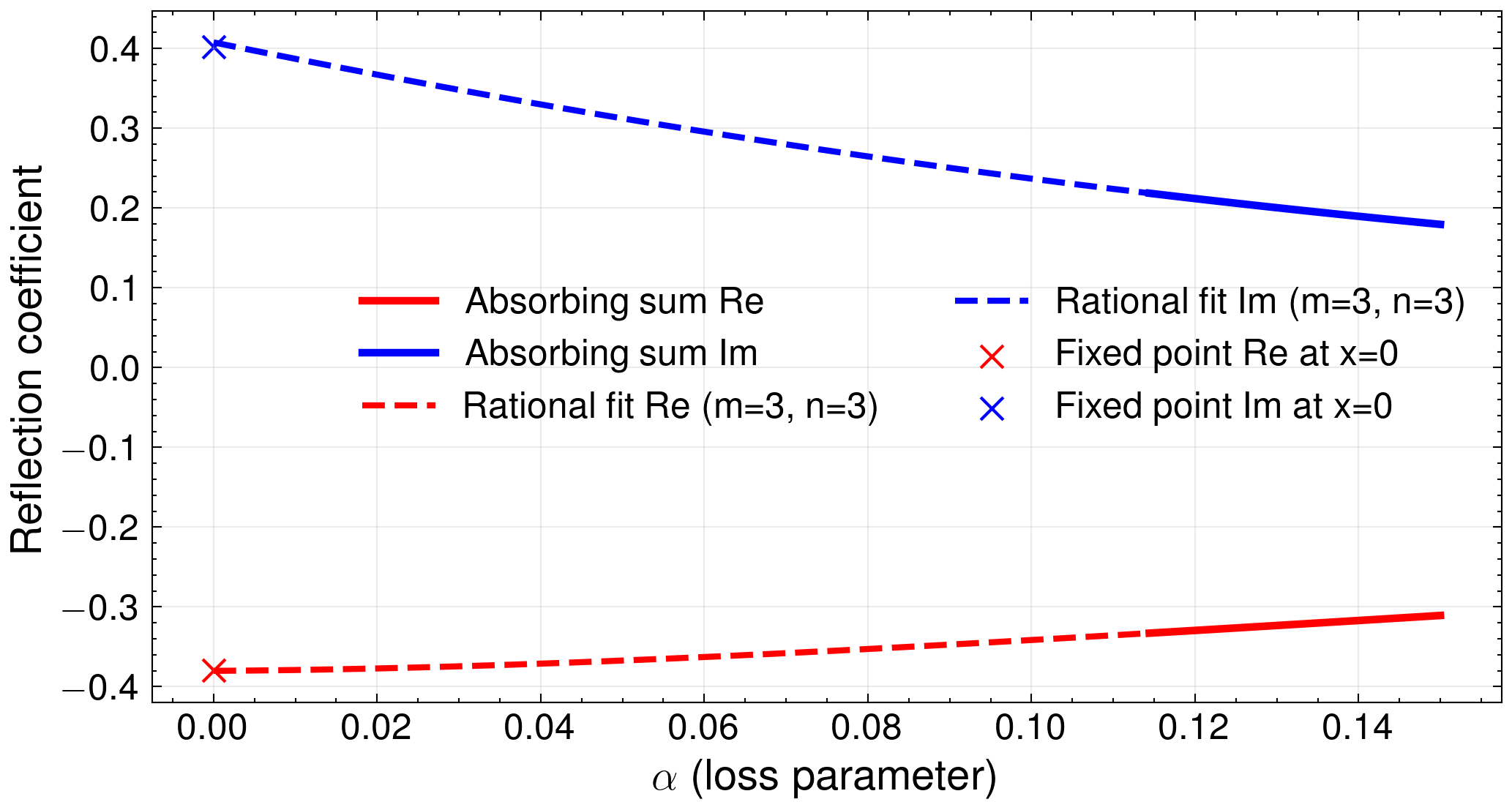}
\caption{
Reflected field coefficient for the structure discussed in figure (Fig. \ref{fig:2}).
Solid curves: real (red) and imaginary (blue) parts of the Neumann sum for decreasing absorption $\alpha$.
Dashed curves: rational fit extrapolating to $\alpha \to 0^+$.
Crosses: fixed-point solution for $\alpha=0$.
}
\label{fig:absorption_fit}
\end{figure}

Figure~\ref{fig:absorption_fit} shows that the introduction of a small imaginary part in the refractive indices provides regularisation that ensures convergence of the multiple-scatter expansion.  
As absorption decreases, the truncated series approaches the divergent regime, but the rational fit captures the limiting behaviour, enabling the estimate of the lossless response. 

\section{Divergence in birefringent materials}

In the case of anisotropic layers, we can express the propagation in the $\bf \hat n$ direction of a plane wave is modelled using the decomposition of Maxwell's equations into perpendicular ($E_z \bf \hat n$, $H_z \bf \hat n$) and transversal (interface parallel) field components ($\bf E^\perp$, $\bf H^\perp$). Similarly,  the wavevector decomposes to ${\bf k}={\bf k^\perp}+k_n {\bf \hat n}$. In this case, we have:
\begin{eqnarray}
k_n {\bf \hat n}\times {\bf E^\perp} &=& \mu_0 \omega {\bf H^\perp}-{\bf k^\perp}\times{\bf \hat n} E_n \nonumber \\
&=&{\bf k^\perp}\times{\bf \hat n} \frac{({\bf \hat n}\cdot\tilde{\boldsymbol\varepsilon}\cdot {\bf E^\perp})}{({\bf \hat n}\cdot\tilde{\boldsymbol\varepsilon}\cdot {\bf \hat n})} +{\bf k^\perp}\times{\bf \hat n} 
\frac{{\bf \hat n}\cdot ({\bf k^\perp}\times {\bf H^\perp})}{\omega({\bf \hat n}\cdot\tilde{\boldsymbol\varepsilon}\cdot {\bf \hat n})} +\mu_0 \omega {\bf H^\perp} \label{ephp1}\\
k_n {\bf \hat n}\times {\bf H^\perp} &=& 
-\omega \tilde{\boldsymbol\varepsilon}\cdot {\bf E^\perp}-\omega \tilde{\boldsymbol\varepsilon}\cdot {\bf \hat n} E_n-({\bf k^\perp}\times {\bf H^\perp})-{\bf k^\perp}\times{\bf \hat n} H_n \nonumber \\
&=&\left( \frac{\tilde{\boldsymbol\varepsilon}\cdot {\bf \hat n}}{({\bf \hat n}\cdot\tilde{\boldsymbol\varepsilon}\cdot {\bf \hat n})}-{\bf \hat n}\right) {\bf \hat n}\cdot ({\bf k^\perp}\times {\bf H^\perp})-
{\bf k^\perp}\times{\bf \hat n}  \frac{{\bf \hat n}\cdot ({\bf k^\perp}\times {\bf E^\perp})}{\omega \mu_0} \\
&&-\omega \tilde{\boldsymbol\varepsilon}\cdot {\bf E^\perp}+\tilde{\boldsymbol\varepsilon}\cdot{\bf \hat n} 
\frac{\omega({\bf \hat n}\cdot\tilde{\boldsymbol\varepsilon}\cdot {\bf E^\perp})}{({\bf \hat n}\cdot\tilde{\boldsymbol\varepsilon}\cdot {\bf \hat n})} \label{ephp2}
\end{eqnarray}
where we used the  perpendicular field components defined by: 
\begin{eqnarray}
E_n &=& -\frac{\omega({\bf \hat n}\cdot\tilde{\boldsymbol\varepsilon}\cdot {\bf E^\perp}) +{\bf \hat n}\cdot({\bf k^\perp}\times {\bf H^\perp})}{\omega ({\bf \hat n}\cdot\tilde{\boldsymbol\varepsilon}\cdot {\bf \hat n})} \\
H_n &=& \frac{{\bf \hat n}\cdot({\bf k^\perp}\times {\bf E^\perp})}{\omega\mu_0}
\label{eq:ezhz}
\end{eqnarray}

Equations (\ref{ephp1}) and (\ref{ephp2}) can be solved as a generalised eigenvalue problems where $k^{(i)}_n$ are the four eigenvalues and  ($\bf E_i^\perp$, $\bf H_i^\perp$) the four eigenvectors representing the four corresponding polarisations. 
The plane wave solutions are: 
\begin{eqnarray}
{\bf E}_i({\bf r},t) &=& ({\bf E}_i^\perp+E_n^{(i)}{\bf \hat n})\exp(\imath\,({\bf k}\cdot {\bf r}-\omega t)) \\
{\bf H}_i({\bf r},t) &=& ({\bf H}_i^\perp+H_n^{(i)}{\bf \hat n})\exp(\imath\,({\bf k}\cdot {\bf r}-\omega t)) 
\end{eqnarray}

In the following, we consider propagation in the z-direction and define the transversal field vector $\bm \Phi=(E_x,E_y,H_x,H_y)^T$ with all the components of the transverse field. In this case, the Berreman first-order system for anisotropic media reads
\begin{equation}
\frac{d}{dz}\,\bm \Phi = \imath\,\mathbf B\,\bm \Phi,
\label{eq:BerremanEQ}
\end{equation}
where the $4\times4$ Berreman matrix $\mathbf B$ is
\begin{equation}
\mathbf B =
\left(
\begin{array}{cccc}
- \dfrac{k_x \tilde{\varepsilon}_{zx}}{\tilde{\varepsilon}_{zz}} &
- \dfrac{k_x \tilde{\varepsilon}_{zy}}{\tilde{\varepsilon}_{zz}} &
\dfrac{k_x k_y}{\omega \tilde{\varepsilon}_{zz}} &
\omega \mu_0 - \dfrac{k_x^2}{\omega \tilde{\varepsilon}_{zz}} \\[8pt]
- \dfrac{k_y \tilde{\varepsilon}_{zx}}{\tilde{\varepsilon}_{zz}} &
- \dfrac{k_y \tilde{\varepsilon}_{zy}}{\tilde{\varepsilon}_{zz}} &
\dfrac{k_y^2}{\omega \tilde{\varepsilon}_{zz}} - \omega \mu_0 &
- \dfrac{k_x k_y}{\omega \tilde{\varepsilon}_{zz}} \\[8pt]
- \dfrac{k_x k_y}{\omega \mu_0} - \omega \tilde{\varepsilon}_{yx} + \omega \dfrac{\tilde{\varepsilon}_{yz}\tilde{\varepsilon}_{zx}}{\tilde{\varepsilon}_{zz}} &
\dfrac{k_x^2}{\omega \mu_0} - \omega \tilde{\varepsilon}_{yy} + \omega \dfrac{\tilde{\varepsilon}_{yz}\tilde{\varepsilon}_{zy}}{\tilde{\varepsilon}_{zz}} &
- \dfrac{\tilde{\varepsilon}_{yz} k_y}{\tilde{\varepsilon}_{zz}} &
\dfrac{\tilde{\varepsilon}_{yz} k_x}{\tilde{\varepsilon}_{zz}} \\[8pt]
- \dfrac{k_y^2}{\omega \mu_0} + \omega \tilde{\varepsilon}_{xx} - \omega \dfrac{\tilde{\varepsilon}_{xz}\tilde{\varepsilon}_{zx}}{\tilde{\varepsilon}_{zz}} &
\dfrac{k_x k_y}{\omega \mu_0} + \omega \tilde{\varepsilon}_{xy} - \omega \dfrac{\tilde{\varepsilon}_{xz}\tilde{\varepsilon}_{zy}}{\tilde{\varepsilon}_{zz}} &
\dfrac{\tilde{\varepsilon}_{xz} k_y}{\tilde{\varepsilon}_{zz}} &
- \dfrac{\tilde{\varepsilon}_{xz} k_x}{\tilde{\varepsilon}_{zz}}
\end{array}
\right).
\label{eq:BerremanMatrix}
\end{equation}

\noindent
For a homogeneous layer (constant $\tilde{\boldsymbol\varepsilon}$ and fixed transverse wavevector components $k_x$, $k_y$), solutions of the form
\(\bm \Phi(z) = \bm \Phi_i\,e^{\imath\, k_z^{(i)} z}\) lead to the generalised eigenvalue problem
\begin{equation}
\mathbf B\,\bm \Phi_i = k_z^{(i)}\,\bm \Phi_i,
\end{equation}
whose four eigenvalues \(k_z^{(i)}\) give the longitudinal wavevector components and whose eigenvectors \(\bm \Phi_i=(E_x^{(i)},E_y^{(i)},H_x^{(i)},H_y^{(i)})^{\mathsf T}\) determine the associated polarisations. The full fields then follow from
\begin{eqnarray}
{\bf E}_i({\bf r},t) &=& \big(E_x^{(i)}\,\hat{\bf x} + E_y^{(i)}\,\hat{\bf y} + E_z^{(i)}\,\hat{\bf z}\big)\,\exp\big(\imath\,({\bf k}\cdot{\bf r}-\omega t)\big),\\
{\bf H}_i({\bf r},t) &=& \big(H_x^{(i)}\,\hat{\bf x} + H_y^{(i)}\,\hat{\bf y} + H_z^{(i)}\,\hat{\bf z}\big)\,\exp\big(\imath\,({\bf k}\cdot{\bf r}-\omega t)\big),
\end{eqnarray}
with \(E_z^{(i)}\) and \(H_z^{(i)}\) given by the elimination formulas above.

We define the quadratic form for the
time-averaged Poynting flux in the z direction as
\[
\langle S_z\rangle=\bm\Phi^\dagger \bm P_z\,\bm\Phi,\qquad
\mathbf P_z=\frac{1}{4}
\begin{pmatrix}
0&0&0&1\\
0&0&-1&0\\
0&-1&0&0\\
1&0&0&0
\end{pmatrix},
\]
so that \(\langle S_z\rangle=\tfrac12\mathrm{Re}(E_x H_y^*-E_y H_x^*)\) while the reactive power is defined by \(\langle Q_z\rangle=\mathrm{Im}(E_x H_y^*-E_y H_x^*)\).

Differentiating the flux and using \eqref{eq:BerremanMatrix} gives
\begin{align}
\frac{\partial}{\partial z}\langle S_z\rangle
&= \frac{\partial}{\partial z}\big(\bm\Phi^\dagger \mathbf P_z\,\bm\Phi\big)
= \big(\tfrac{\partial\bm\Phi}{\partial z}\big)^\dagger \mathbf P_z\,\bm\Phi
   + \bm\Phi^\dagger \mathbf P_z\,\big(\tfrac{\partial\bm\Phi}{\partial z}\big) \nonumber\\
&= (-\imath\,\bm\Phi^\dagger \mathbf B^\dagger)\,\bm P_z\,\bm\Phi
   + \bm\Phi^\dagger \mathbf P_z\,(\imath\,\mathbf B\,\bm\Phi) \nonumber\\
&= \imath\,\bm\Phi^\dagger\big(\mathbf P_z\mathbf B-\mathbf B^\dagger \mathbf P_z\big)\bm\Phi.
\label{eq:dz_flux_general}
\end{align}
Since for a source-free and lossless medium (\(\tilde{\boldsymbol\varepsilon}=\tilde{\boldsymbol\varepsilon}^\dagger\)), the  Poynting theorem yields \(\partial_z \langle S_z\rangle=0\), we obtain the algebraic condition
\begin{equation}
\mathbf P_z\mathbf B=\mathbf B^\dagger \mathbf P_z
\label{eq:P_hermiticity}
\end{equation}
which corresponds to the \(\mathbf P_z\)-Hermiticity of the Berreman operator.

We define the interference time-averaged Poynting flux between two transverse fields \(\bm \Phi_i\) and \(\bm \Phi_j\) as
$$\langle S_z\rangle_{ij}=\bm\Phi_i^\dagger \bm P_z\,\bm\Phi_j. $$
Let \(\bm \Phi_i\) and \(\bm \Phi_j\) be eigenvectors of $\bf B$. In this case we have:
$$
k_z^{(j)}\bm\Phi_i^\dagger \bm P_z\,\bm\Phi_j=\bm\Phi_i^\dagger \bm P_z\,\bm B\,\bm\Phi_j=\bm\Phi_i^\dagger \bm B^\dagger\,\bm P_z\,\bm\Phi_j = k_z^{(i)\ast}\bm\Phi_i^\dagger \bm P_z\,\bm\Phi_j
$$
Implying
\begin{equation}
(k_z^{(j)}-k_z^{(i)\ast})(\bm\Phi_i^\dagger \bm P_z\,\bm\Phi_j) = 0.
\label{eq:orthocondition}
\end{equation}
Therefore, if \(k_z^{(j)}\neq k_z^{(i)\ast}\), the eigenvectors are orthogonal in the flux metric and do not interfere with respect to the power measure. In the special case of a complex-conjugate pair \(k_z^{(j)}=k_z^{(i)\ast}\) with
\(\mathrm{Im}\big(k_z^{(i)}\big)\neq 0\), \eqref{eq:orthocondition} imposes no constraint on
\((\bm\Phi_i^\dagger \bm P_z\,\bm\Phi_j)\) which can be non-zero. Moreover, taking \(j=i\) gives
\begin{equation}
(k_z^{(i)}-k_z^{(i)\ast})(\bm\Phi_i^\dagger \bm P_z\,\bm\Phi_i)=0
\quad\Rightarrow\quad
\mathrm{Im}\big(k_z^{(i)}\big)\neq 0 \quad\Rightarrow\quad(\bm\Phi_i^\dagger \bm P_z\,\bm\Phi_i)=0.
\end{equation}
Consequently, each eigenvector in a non-real conjugate pair has zero power flux, while their interference flux
is generally non-zero.

For each non-real conjugate pair $(i,j)$, we can define two power modes: 
$$\frac{1}{\sqrt{2}}\left( \frac{\Phi_{i}}{\sqrt{<S_z>_{ij}}}\pm\frac{\Phi_{j}}{\sqrt{<S_z>_{ji}}}\right)$$
corresponding to two eigenmodes of $\bm P_z$ with eigenvalues $\pm 1$.
Furthermore, it is possible to use each eigenmode of $\bm B$, regardless of its pairing and eigenvalue, to define associated power eigenmodes by introducing $\chi_i \equiv \operatorname{sgn}\!\big(\operatorname{Im} \big(k^{(i)}_z\big)\big)\in\{-1,0,1\}$.  The associated power eigenmode is: 
\begin{equation}
\bm \Psi_i(z)=    
\frac{1}{\sqrt{1+\chi^2_i}}
\left( 
\frac{\bm\Phi_{i} }{\sqrt{\big(\bm \Phi_i^T \bm P_z \bm\Phi_i\big)}}\exp\left(\imath\,k^{(i)}_z z\right)
+\chi_i\frac{\bm\Phi^\ast_{i}}{\sqrt{\big(\bm \Phi_i^T \bm P_z \bm\Phi_i\big)^\ast}}\exp\left(\imath\,k^{(i)\ast}_z z\right)
\right).
\label{eq:powermode}
\end{equation}
Note that the normalisation denominator terms only involve the transpose of the transverse field vector $\bm \Phi$ without conjugation. This is important to remove the interference terms and normalise power flux. 

This allows us to define the energy velocity in the z-direction as:
\[
v_{E,z}=\frac{\langle S_z\rangle}{u}.
\]
where the energy density is defined by:
\[
u=\frac{1}{4}\left[\mathbf E^\dagger\cdot\tilde{\boldsymbol\varepsilon}\cdot\mathbf E
+
\mu_0\mathbf H^\dagger\cdot\mathbf H\right].
\]
where $E_z$ and $H_z$ are constructed using (\ref{eq:ezhz}).

\begin{figure}
    \centering
    \includegraphics[width=1.0\linewidth]{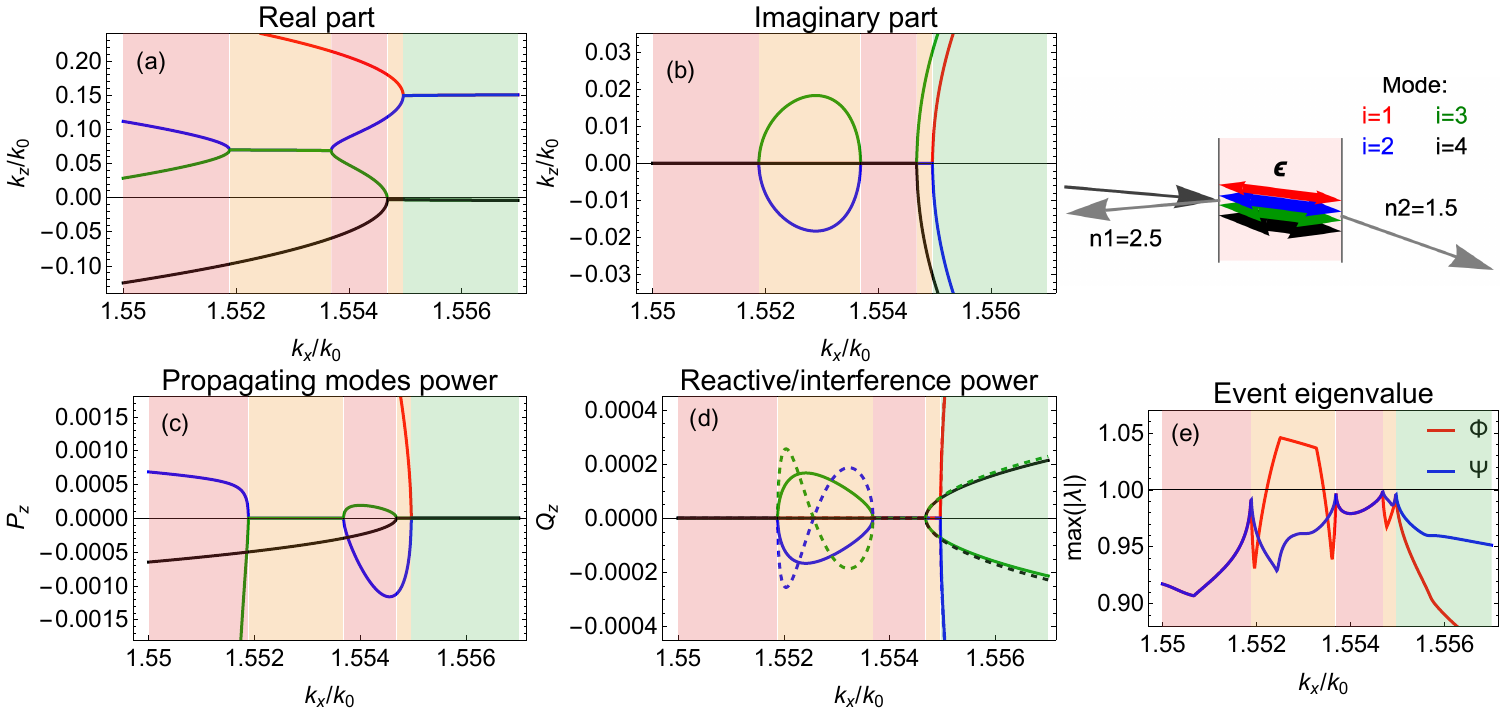}
    \caption{Real (a) and imaginary (b) parts of the propagation constant for the eigemodes of the propagation matrix $\bf B$. (c) Transported power flux and  (d) reactive (plain) and interference (dashed) power of the eigenmodes of $\bf B$. (e) Maximal event scattering eigenvalue for the propagation matrix eigenmodes (red) and for the power modes (blue). Values above one correspond to divergent series. Background colouring highlights the different regimes with respect to the critical angles: light red corresponds to media having four propagating eigenmodes, in the light yellow region we have two propagating and two evanescent modes and light green corresponds to four evanescent modes.  }
    \label{fig:sup:anikz}
\end{figure}

\begin{figure}
    \centering
    \includegraphics[width=0.95\linewidth]{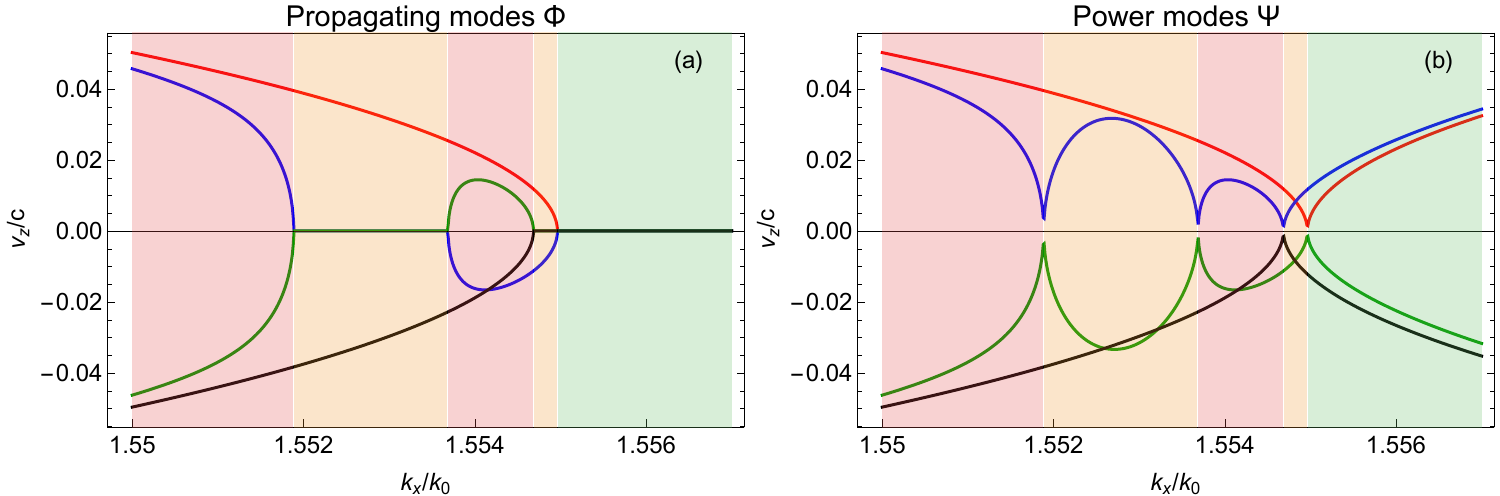}
    \caption{Energy velocity as a function of $k_x$ wavevector for (a) the propagating eigenmodes and for (b) the power modes. Mode colouring is the same as in Figure \ref{fig:sup:anikz}.}
    \label{fig:Ve}
\end{figure}

As a numerical example, we consider a thin ($h=20$~nm) layer of Whewellite  anisotropic crystal ($n_\alpha=1.491$, $n_\beta=1.554$, $n_\gamma=1.650$) at a wavelength of 620 nm. The input and output semi-spaces are TiO2 ($n=2.49$) and SiO2 ($n=1.458$) respectively. We rotated the crystal using the Euler angles (using the XYX convention, 70 deg, 51 deg, 88 deg). The angle of the crystal is chosen so that the varying incident angle in the $y=0$ plane scans across two critical angles and the mode mixing region, leading to divergence of the event matrix. Figures \ref{fig:sup:anikz} and \ref{fig:Ve} show that the series of multiple scattering events diverges in a single anisotropic layer when the two slow modes mix in the evanescent regime. Figure \ref{fig:Ve}b shows that the power modes remove the mixing degeneracy which maintains the expected power flux direction and causality in scattering events. 

\section{Divergence in elastic media}

Let $\rho$ be the mass density and $C_{ijkl}$ the fourth-order stiffness tensor. The equations of motion and the constitutive law are
\[
\rho\,\partial^2_t {u}_i = \partial_j\sigma_{ij}, \qquad
\sigma_{ij} = C_{ijkl}\,\varepsilon_{kl}, \qquad
\varepsilon_{kl} = \tfrac{1}{2}\,(\partial_lu_{k}+\partial_ku_{l}),
\]
where $u$ is displacement, $\bm \varepsilon$ strain, and $\bm \sigma$ stress. We consider time-harmonic fields with fixed wavevector components in the $(x,y)$ plane $e^{\imath\, k_x x - \imath\, \omega t}$. 
We introduce the $3\times 3$ matrices that gather the material response along $z$:
\[
{\bm\Gamma}_\mathbf{ik} \in \mathbb{R}^{3\times 3}, \quad (\mathbf{\bm \Gamma_{ik}})_{jl} = C_{jilk},
\]
We define the velocity field $v_i=\partial_t u_i$ so that, $v_i=-\imath\,\omega u_i$ and the normal stress $\sigma_{i3}=\bm\sigma\,\mathbf{e}_3$. One obtains the Stroh/Adler system for the state vector $\bm \phi=(v_1,v_2,v_3,\sigma_{13},\sigma_{23},\sigma_{33})$:
\begin{equation}
\frac{d \bm \phi}{dz}
= \imath\,\mathbf{A}\bm \phi,
\qquad
\mathbf{A} \;=\;
\begin{bmatrix}
-k_x\mathbf{X}\mathbf{\bm \Gamma_{31}} & -\,\omega\,\mathbf{X} \\
\ \dfrac{1}{\omega}\Big(k_x^2\mathbf{\bm \Gamma_{11}} - \rho\,\omega^{2}\mathbf{I} - k_x^2\mathbf{\bm \Gamma_{13}}\mathbf{X}\mathbf{\bm \Gamma_{31}}\Big) & -\,k_x\mathbf{\bm \Gamma_{13}}\mathbf{X}
\end{bmatrix}.
\label{eq:stroh-vt}
\end{equation}
where $\bf X={\bm \Gamma^{-1}_\mathbf{33}}$.
In a lossless medium, $\mathbf{A}$ is $\mathbf{J}$-Hermitian, i.e.
\[
\mathbf{A}^\dagger \mathbf{J} = \mathbf{J}\,\mathbf{A},
\qquad
\mathbf{J} \;=\; \begin{bmatrix} \mathbf{0} & -\,\mathbf{I} \\ -\,\mathbf{I} & \mathbf{0} \end{bmatrix}.
\]
This implies the conservation of 
\[
\frac{d}{dz}\left({\bm\phi}^{\dagger}\mathbf{J}{\bm\phi}\right)=0.
\]
The instantaneous mechanical power flux (Umov–Poynting vector) is $\mathbf{S} = -\,\bm\sigma\,\bf v$. Its $z$-component time-average over one cycle is
\[
\langle S_z \rangle \;=\; \tfrac{1}{4}\,{\bm \phi}^{\dagger}\mathbf{J}{\bm\phi} 
\]

The total mechanical energy density (kinetic plus strain) averaged over one period is:
\[
\langle e \rangle \;=\; \tfrac{1}{4}\,\rho {\bf v}^{\ast}\, {\bf v}+\tfrac{1}{4}\,{\bm \sigma}^\ast:{\bm \varepsilon}
\]
defining the energy velocity in the z-direction as:
\[
v_{E,z}=\frac{\langle S_z\rangle}{\langle e \rangle}.
\]

In the following, we consider the isotropic medium defined by: $C_{ijkl}=\lambda\,\delta_{ij}\,\delta_{kl}
+ \mu\left(\delta_{ik}\,\delta_{jl} + \delta_{il}\,\delta_{jk}\right)$. 
The six eigenvector $\bm\phi_i=(\bm v_i\, ,\,\bm\sigma_i\cdot\bm n)$ with $\bm n=\bm e_3$ of $\bf A$ are:

\[
\begin{aligned}
\mathbf{v}_1 &= \frac{1}{\sqrt{k_{x}^{2}+ k_{zS}^{2}}}\left(  k_{zS},\ 0,\ -k_{x} \right), &
(\bm \sigma_{1}\!\cdot\! \bm n) &= \frac{1}{\sqrt{k_{x}^{2}+ k_{zS}^{2}}} \left( \frac{(2k_{x}^{2}\mu-\rho\,\omega^{2})}{\omega},\ 0,\ \frac{2k_{x}\mu k_{zS}}{\omega} \right),\\[6pt]
\mathbf{v}_2 &= \left( 0,\ 1,\ 0 \right), &
(\bm \sigma_{2}\!\cdot\! \bm n) &= \left( 0,\ -\frac{\mu\, k_{zS}}{\omega},\ 0 \right),\\[6pt]
\mathbf{v}_3 &= \frac{k_{x}}{\sqrt{k_{x}^{2}+ k_{zL}^{2}}}\left( k_{x},\ 0,\ k_{zL} \right), &
(\bm \sigma_{3}\!\cdot\! \bm n) &= \frac{k_{x}}{\sqrt{k_{x}^{2}+ k_{zL}^{2}}}\left( -\frac{2k_{x}k_{zL}\mu}{\omega},\ 0,\ -\frac{k_{x}^{2}\lambda+k_{zL}(\lambda+2\mu)}{\omega} \right),\\[6pt]
\mathbf{v}_4 &= \frac{1}{\sqrt{k_{x}^{2}+ k_{zS}^{2}}}\left( k_{zS},\ 0,\ k_{x} \right), &
(\bm \sigma_{4}\!\cdot\! \bm n) &= \frac{1}{\sqrt{k_{x}^{2}+ k_{zS}^{2}}}\left( \frac{(-2k_{x}^{2}\mu+\rho\,\omega^{2})}{\omega},\ 0,\ \frac{2k_{x}\mu k_{zS}}{\omega} \right),\\[6pt]
\mathbf{v}_5 &= \left( 0,\ 1,\ 0 \right), &
(\bm\sigma_{5}\!\cdot\! \bm n) &= \left( 0,\ \frac{\mu\, k_{zS}}{\omega},\ 0 \right),\\[6pt]
\mathbf{v}_6 &= \frac{k_{x}}{\sqrt{k_{x}^{2}+ k_{zL}^{2}}}\left( k_{x},\ 0,\ -k_{zL} \right), &
(\bm \sigma_{6}\!\cdot\! \bm n) &= \frac{k_{x}}{\sqrt{k_{x}^{2}+ k_{zL}^{2}}}\left( \frac{2k_{x}k_{zL}\mu}{\omega},\ 0,\ -\frac{k_{x}^{2}\lambda+k_{zL}(\lambda+2\mu)}{\omega} \right)
\end{aligned}
\]
with the corresponding eigenvalues $k^{(i)}_{z}$=\Big($k_{zS},\,k_{zS},\,k_{zL},\,-k_{zS},\,-k_{zS},\,-k_{zL}$\Big) where $k_{zS}=\sqrt{\frac{\rho}{\mu}\,\omega^{2} - k_{x}^{2}}$  and $k_{zL}= \sqrt{\frac{\rho}{\lambda + 2\,\mu}\,\omega^{2} - k_{x}^{2}}$.

\vspace*{5mm}
\hspace*{-30mm}
{\small\(
\frac{1}{4}\bm \phi_i^\ast{\bf J}\bm\phi_j=\begin{pmatrix}
\dfrac{\mu \left(k_{zS}+k_{zS}^{\ast}\right)}{4\omega} & 0 & 0 &
\dfrac{\mu\left(k_{zS}-k_{zS}^{\ast}\right)}{4\omega} & 0 & 0
\\[6pt]
0 & \dfrac{\mu\left(k_{zS}+k_{zS}^{\ast}\right)}{4\omega} & 0 &
0 & -\dfrac{\mu\left(k_{zS}-k_{zS}^{\ast}\right)}{4\omega} & 0
\\[6pt]
0 & 0 & \dfrac{(\lambda+2\mu)\left(k_{zL}+k_{zL}^{\ast}\right)}{4\omega} &
0 & 0 & -\dfrac{(\lambda+2\mu)\left(k_{zL}-k_{zL}^{\ast}\right)}{4 \omega}
\\[6pt]
-\dfrac{\mu\left(k_{zS}-k_{zS}^{\ast}\right)}{4\omega} & 0 & 0 &
-\dfrac{\mu\left(k_{zS}+k_{zS}^{\ast}\right)}{4 \omega} & 0 & 0
\\[6pt]
0 & \dfrac{\mu\left(k_{zS}-k_{zS}^{\ast}\right)}{4 \omega} & 0 &
 0 & -\dfrac{\mu\left(k_{zS}+k_{zS}^{\ast}\right)}{4 \omega} & 0
\\[6pt]
0 & 0 & \dfrac{(\lambda+2 \mu)\left(k_{zL}-k_{zL}^{\ast}\right)}{4\omega} &
0 & 0 & -\dfrac{(\lambda+2\mu)\left(k_{zL}+k_{zL}^{\ast}\right)}{4\omega}
\end{pmatrix}_{ij}
\)}

\vspace*{5mm}
Defining the power modes using the function $\chi_i \equiv \operatorname{sgn}\!\big(\operatorname{Im} \big(k^{(i)}_z\big)\big)\in\{-1,0,1\}$

\begin{equation}
\bm \psi_i(z)=    
\frac{2}{\sqrt{1+\chi^2_i}}
\left( 
\frac{\bm\phi_{i} }{\sqrt{\big(\bm \phi_i \bf J \bm\phi_i\big)}}\exp\left(\imath\,k^{(i)}_z z\right)
+\chi_i\frac{\bm\phi^\ast_{i}}{\sqrt{\big(\bm \phi_i \bf J \bm\phi_i\big)^\ast}}\exp\left(\imath\,k^{(i)\ast}_z z\right)
\right).
\label{eq:powermode2}
\end{equation}

with
\[
\bm \psi^\ast_i{\bf J}{\bm\psi}_j=\begin{pmatrix}
1 & 0 & 0 & 0 & 0 & 0 \\[6pt]
0 & 1& 0 & 0 & 0 & 0 \\[6pt]
0 & 0 & 1 & 0 & 0 & 0 \\[6pt]
0 & 0 & 0 &-1 & 0 & 0 \\[6pt]
0 & 0 & 0 & 0 & -1 & 0 \\[6pt]
0 & 0 & 0 & 0 & 0 & -1 
\end{pmatrix}_{ij}
\]

\begin{figure}
    \centering
    \includegraphics[width=1.0\linewidth]{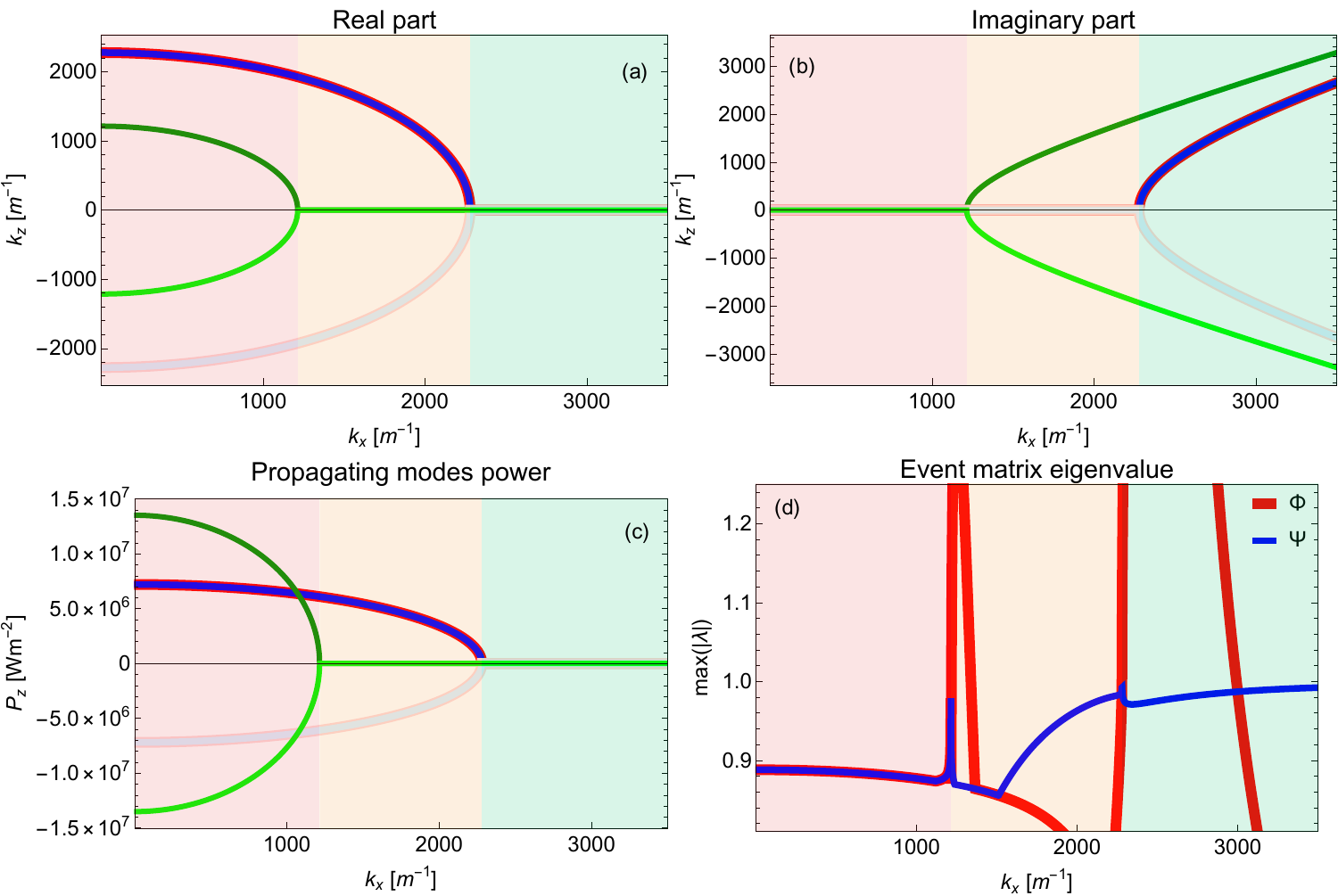}
    \caption{Real (a) and imaginary (b) parts of the propagation constant for the eigemodes of the propagation matrix $\bf A$.  (c) Transported power flux of the eigenmodes of $\bf A$. The eigenmodes are normalised with respect to the associated velocity. (d) Maximal event scattering eigenvalue for the propagation matrix eigenmodes (red) and for the power modes (blue). Values above one correspond to divergent series. Background colouring highlights the different regimes with respect to the critical angles: light red corresponds to media having four propagating eigenmodes, in the light yellow region we have two propagating and two evanescent modes and light green corresponds to four evanescent modes.  }
    \label{fig:sup:USiso}
\end{figure}

\begin{figure}
    \centering
    \includegraphics[width=0.95\linewidth]{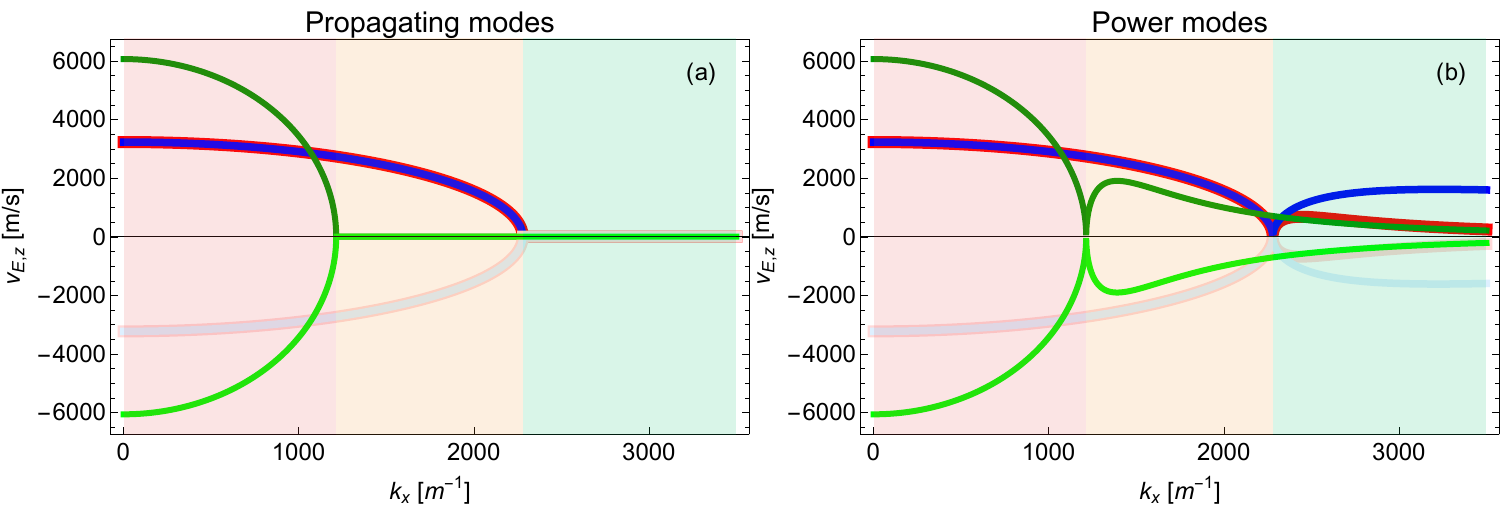}
    \caption{Energy velocity as a function of $k_x$ wavevector for (a) the propagating eigenmodes and for (b) the power modes. Mode colouring is the same as in Figure \ref{fig:sup:USiso}.}
    \label{fig:USVe}
\end{figure}

\begin{figure}
    \centering
    \includegraphics[width=1.0\linewidth]{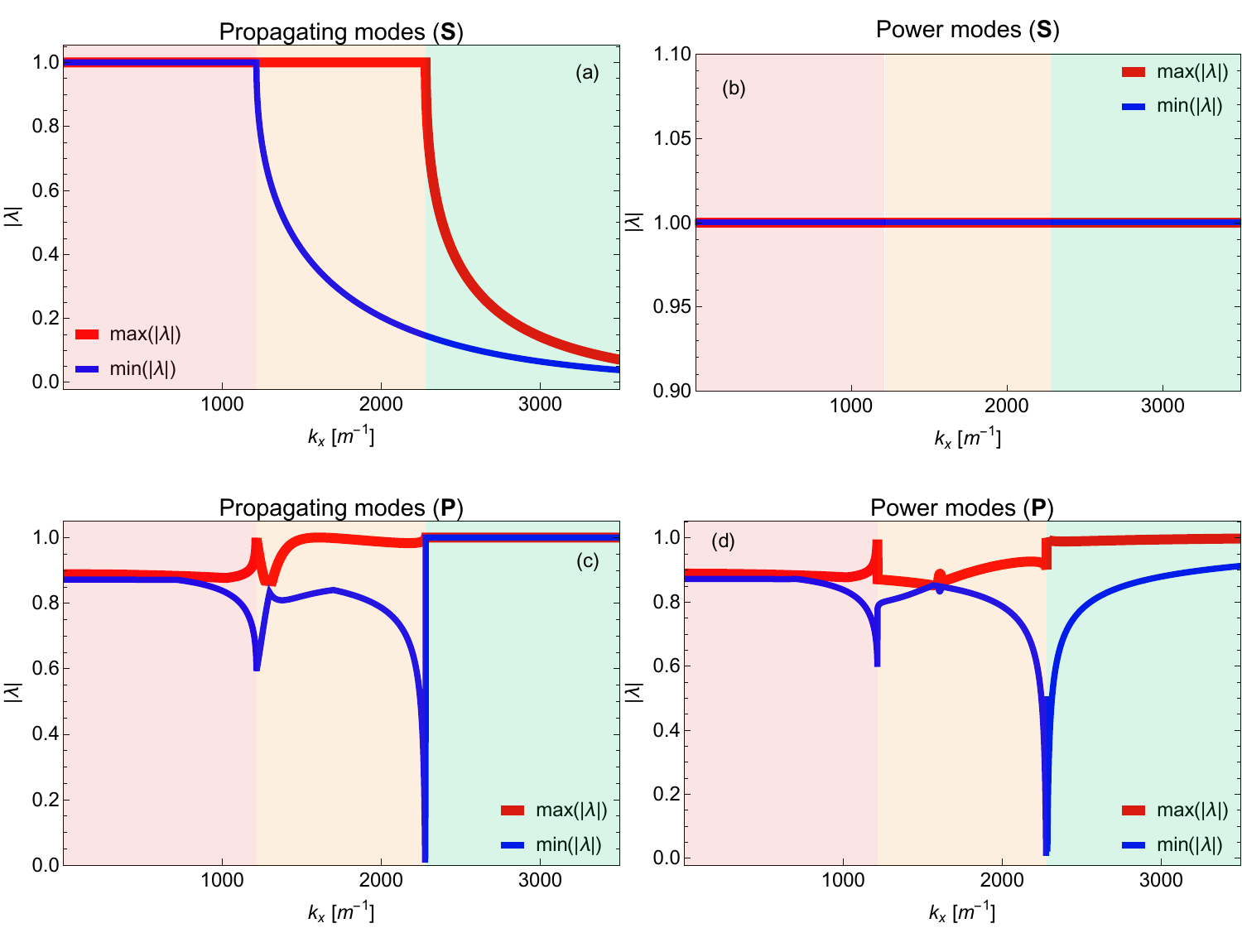}
    \caption{Max and min magnitudes of the scattering and propagation eigenvalues for the propagation matrix eigenmodes (a,c) and for the power modes (b,d).  }
    \label{fig:sup:USisoSP}
\end{figure}

As a numerical example, we consider a thin ($h=1$~mm, medium (2)) layer of titanium ($v_{L}=\sqrt{\frac{\lambda + 2\,\mu}{\rho}}=6060$~m/s, $v_S=\sqrt{\frac{\mu}{\rho}}=3230$~m/s, $\rho=4460$~kg/m$^3$) at a frequency of 7.36~MHz. The input and output semi-spaces (medium (1)) are polyethylene ($v_L=1950$~m/s, $v_S=900$~m/s, $\rho=950$~kg/m$^3$).  Figures \ref{fig:sup:USiso} and \ref{fig:USVe} show that the series of multiple scattering events diverges in the  layer when the modes become evanescent. Figure \ref{fig:USVe}b shows that the power modes maintain the expected power flux direction and causality in scattering events even when evanescent. Figure \ref{fig:sup:USisoSP} shows that the power mode scattering matrix $\bf S$ is unitary while the propagation matrix $\bf P$ has eigenvalues bound by the unit circle. 

\begin{figure}
    \centering
    \includegraphics[width=1.0\linewidth]{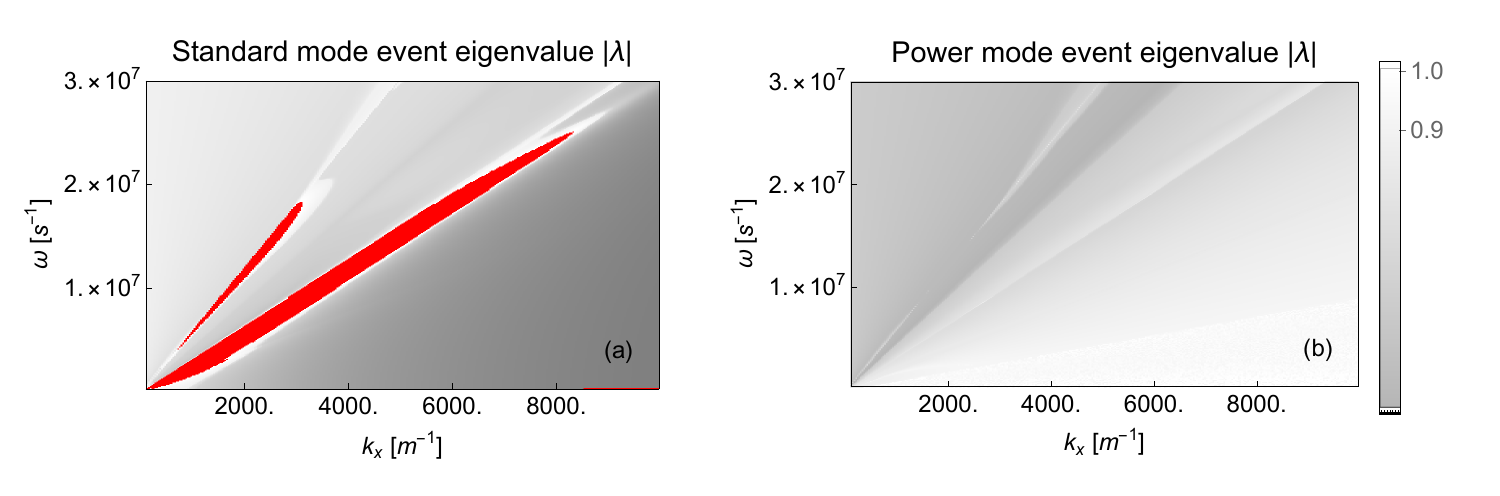}
    \caption{Magnitude of the event scattering matrix  eigenvalues for (a) the standard propagation modes ${\bf P}_s{\bf S}_s$ and for (b) the power modes ${\bf P}_p{\bf S}_p$ as a function of the frequency $\omega$ and horizontal wavevector component $k_x$. The red coloured region indicates eigenvalues larger then one.}
    \label{fig:sup:lamUS}
\end{figure}

Figure~\ref{fig:sup:lamUS} shows the magnitude of the eigenvalues of the event scattering matrix $\mathbf{P}_{s}\mathbf{S}_{s}$. To define the two constituents of the event scattering matrix, it is useful to define the conversion between the transfer matrix and reflection and transmission scattering matrix:
\begin{eqnarray}
\mathbf{S}({\bf M})=\begin{pmatrix}
        \mathbf{R}_{12}({\bf M})& &
        \mathbf{T}_{21}({\bf M}) \\
        \mathbf{T}_{12}({\bf M}) & &
\mathbf{R}_{21}({\bf M})
\end{pmatrix}=\begin{pmatrix}
        -\mathbf{M}_{22}^{-1}\mathbf{M}_{21} &\quad&
        \mathbf{M}_{22}^{-1} \\
        \mathbf{M}_{11} - \mathbf{M}_{12}\mathbf{M}_{22}^{-1}\mathbf{M}_{21} & &
\mathbf{M}_{12}\mathbf{M}_{22}^{-1}
\end{pmatrix}
\end{eqnarray}
with:
\begin{equation}
    \mathbf{M} = \begin{pmatrix}
       \mathbf{M}_{11} & \mathbf{M}_{12} \\
        \mathbf{M}_{21} & \mathbf{M}_{22}
    \end{pmatrix}
\end{equation}
where each of the $\bf M_{ij}$, $\bf R_{ij}$ and $\bf T_{ij}$ block matrices are 3 by 3 describing the linear relationships between SV, SH, and L polarisations.

For a given layer of thickness $h$, the propagation of elastic waves is described by the components of the  wavevector $k^{(i)}_{z}$ defining the standard mode propagation matrix, 
\begin{equation}
    \mathbf{P}_{s} = \mathbf{S}\Big(\mathrm{diag}\big(e^{\imath\, h k^{(1)}_z}, e^{\imath\, h k^{(2)}_z}, \dots, e^{\imath\, h k^{(6)}_z}\big)\Big).
\end{equation}
As we only consider a single layer with two identical interfaces, the standard mode scattering matrix $\mathbf{S}_s$ is obtained using the layer transfer matrix:
\begin{equation}
    \mathbf{M}_s =\Big(\bm \phi^{(2)}_{ij}\Big)^{-1}\Big(\bm \phi^{(1)}_{ij}\Big)
\end{equation}
with $\phi_{ij}$ is the $i$-th element of the standard mode $j$ in medium (1) or (2). The transfer matrix ensures continuity of the velocity and normal stress on the interface between medium (1) and (2).

The standard mode scattering matrix is
\begin{equation}
    \mathbf{S}_s = \begin{pmatrix}
       \mathbf{R}_{21}({\bf M}_s) & 0 \\
        0 & \mathbf{R}_{s21}({\bf M}_s)
    \end{pmatrix}.
\end{equation}
Regions where the maximum eigenvalue magnitude exceeds unity (shown in red) indicate divergence in the iterative solution defined by the series:
\begin{equation}
    \mathbf{R}_s(N)= \mathbf{R}_{12}({\bf M}_s) + \mathbf{T}_{21}({\bf M}_s)\bm\Gamma_{36}\left(\sum_{i=0}^{N} (\mathbf{P}_s\mathbf{S}_s)^i\right)\mathbf{P}_s\bm\Gamma_{63}\mathbf{T}_{12}({\bf M}_s),
\end{equation}
where $\mathbf{T}_{21}({\bf M}_s)$ and $\mathbf{T}_{12}({\bf M}_s)$ are interface transmission matrices for standard modes. The rectangular matrices are defined by:
\[
\bm{\Gamma}_{36} =
\begin{bmatrix}
\mathbf{I}_3 & \mathbf{0}_3
\end{bmatrix}
=
\begin{bmatrix}
1 & 0 & 0 & 0 & 0 & 0 \\
0 & 1 & 0 & 0 & 0 & 0 \\
0 & 0 & 1 & 0 & 0 & 0
\end{bmatrix},
\]
and $\bm\Gamma_{63}=\bm\Gamma_{36}^T$ are used to relate the three external incident or reflected modes to six modes inside the layer.

\begin{figure}
    \centering
    \includegraphics[width=1.0\linewidth]{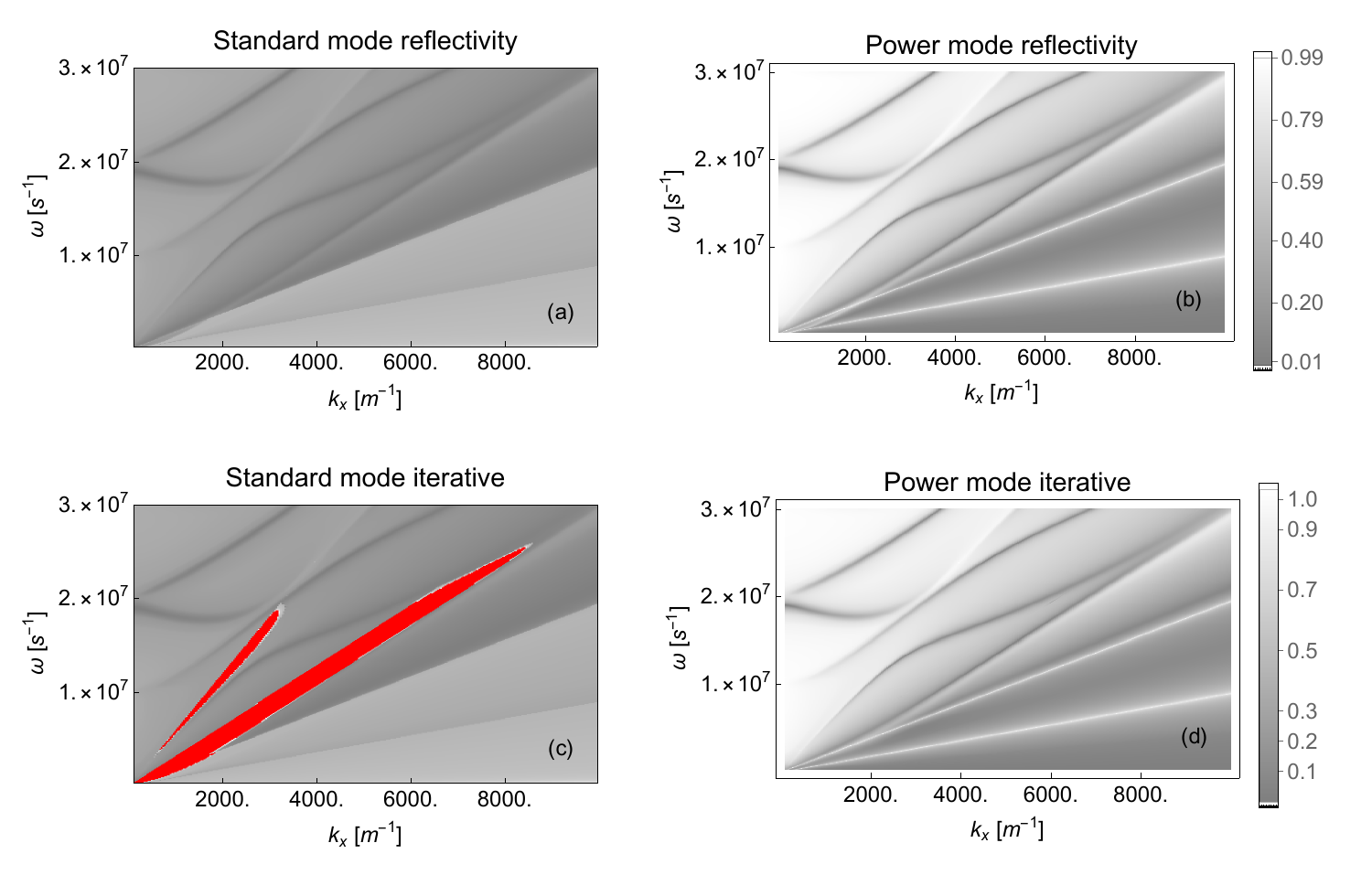}
    \caption{Magnitude of the longitudinal mode layer field reflectivity as a function of the frequency $\omega$ and horizontal wavevector component $k_x$ calculated using four different methods: direct matrix solution for (a) the standard propagation modes $\big({\bf R}_s\big)_{33}$, (b) for the power modes $\big({\bf R}_p\big)_{33}$, iteration method for (c) the standard propagation modes $\big({\bf R}_s(156)\big)_{33}$, and (d) for the power modes $\big({\bf R}_p(156)\big)_{33}$. The red coloured region indicates divergence of the iterative method. }
    \label{fig:sup:refUS}
\end{figure}

The divergence of this series (here highlighted in red) corresponds to eigenvalues of $\mathbf{P}_s\mathbf{S}_s$ that exceed unity, consistent with the eigenvalue analysis in Figure~\ref{fig:sup:lamUS}. For comparison, the closed-form fixed point solution is:
\begin{equation}
    \mathbf{R}_s = \mathbf{R}_{12}({\bf M}_s) + \mathbf{T}_{21}({\bf M}_s)\bm\Gamma_{36}\big(\mathbf{I} - \mathbf{P}_s\mathbf{S}_s\big)^{-1}\mathbf{P}_s\bm\Gamma_{63}\mathbf{T}_{12}({\bf M}_s).
\end{equation}
also shown in Figure~\ref{fig:sup:refUS}. 

Similarly, we define the different matrices using the power modes ($\bm \psi^{(2)}_{i}$) inside the layer by defining the transfer matrix: 
\begin{equation}
    \mathbf{M}_{\rm interface} =\Big(\bm \psi^{(2)}_{ij}\Big)^{-1}\Big(\bm \phi^{(1)}_{ij}\Big)
\end{equation}
which allows us to define the associated reflection and transmission matrices. The power mode scattering matrix is:
\begin{equation}
    \mathbf{S}_p = \begin{pmatrix}
       \mathbf{R}_{21}({\bf M}_{\rm interface}) & 0 \\
        0 & \mathbf{R}_{21}({\bf M}_{\rm interface})
    \end{pmatrix}.
\end{equation}

Inside the layer, the power mode transfer matrix can be expressed using the standard mode transfer matrix as:
\begin{equation}
    \mathbf{M}_p = \Big(\bm \psi^{(2)}_{ij}\Big)^{-1}\Big(\bm \phi^{(2)}_{ij}\Big) \Big(\mathrm{diag}\big(e^{\imath\, h k^{(1)}_z}, e^{\imath\, h k^{(2)}_z}, \dots, e^{\imath\, h k^{(6)}_z}\big)\Big)
    \Big(\bm \phi^{(2)}_{ij}\Big)^{-1}\Big(\bm \psi^{(2)}_{ij}\Big)
\end{equation}
corresponding to a mode basis change from power mode to standard propagating modes and back after the propagation across the layer. This enables the expression of the propagation matrix as:
\begin{equation}
    \mathbf{P}_p = \mathbf{S}\Big(\mathbf{M}_p\Big).
\end{equation}
Using the power modes, the event scattering matrix is $\mathbf{P}_p\mathbf{S}_p$. The iterative reflectivity is defined by:
\begin{equation}
    \mathbf{R}_p(N)= \mathbf{R}_{12}({\bf M}_p) + \mathbf{T}_{21}({\bf M}_p)\bm\Gamma_{36}\left(\sum_{i=0}^{N} (\mathbf{P}_p\mathbf{S}_p)^i\right)\mathbf{P}_p\bm\Gamma_{63}\mathbf{T}_{12}({\bf M}_p),
\end{equation}
and the fixed point matrix solution:
\begin{equation}
    \mathbf{R}_p = \mathbf{R}_{12}({\bf M}_p) + \mathbf{T}_{21}({\bf M}_p)\bm\Gamma_{36}\big(\mathbf{I} - \mathbf{P}_p\mathbf{S}_p\big)^{-1}\mathbf{P}_p\bm\Gamma_{63}\mathbf{T}_{12}({\bf M}_p).
\end{equation}

\section{Input and scattered power flux in elastic waves}

In the previous supplementary section, we used standard modes in the outer medium. This ensures that the solutions are finite at infinity. However, this condition can be relaxed when the fields are observed using transducers. In this case, the input waves need to carry energy, and the power mode representation is necessary to correctly account for the flux, regardless of whether it is propagating or evanescent. 

\begin{figure}
    \centering
    \includegraphics[width=1.0\linewidth]{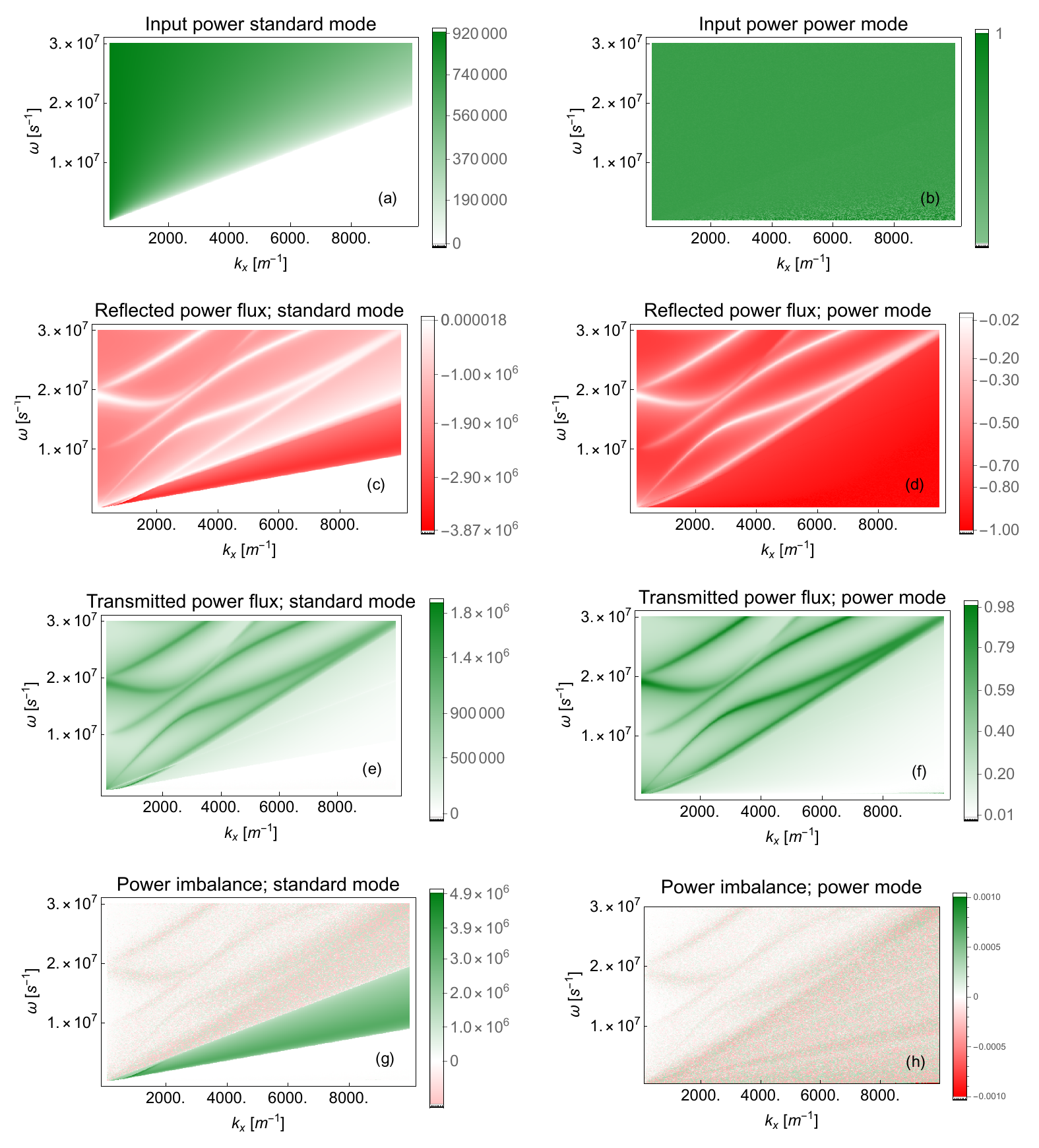}
    \caption{Input, reflected, transmitted and power imbalance from a single layer as a function of the frequency $\omega$ and horizontal wavevector component $k_x$ calculated using standard modes (a,c,e,g)  and power modes (b,d,f,h) in the outer medium. Parts (g,h) show the power flux imbalance $\Delta_z$.}
    \label{fig:sup:powerUS}
\end{figure}

Figure~\ref{fig:sup:powerUS} shows that in the case of standard modes, there is a power imbalance between the incident and scattered field on the incident side of the outer medium. To exemplify this imbalance, we consider a longitudinal wave incident on the single layer
\[
\mathbf{u}_3 \equiv (0,0,1)^{\mathsf T}
\]
where only the third mode is excited with a unit amplitude. This third mode is longitudinal in both representations, standard and power mode; however, it will carry different amounts of power due to the different normalisations employed. We define the respective incident power flux carried by:
\begin{eqnarray}
\langle S_z\rangle_{s}^{(i)} &=& \tfrac{1}{4}\,{\boldsymbol\phi_3}^{\dagger}\,\mathbf{J}\,{\boldsymbol\phi_3},= \tfrac{1}{4}\,\mathbf{u}_3^{\dagger}\,{\boldsymbol\Gamma}_{36}\,\Big(\bm \phi^{(1)}_{ij}\Big)^\dagger\,\mathbf{J}\,\Big(\bm \phi^{(1)}_{ij}\Big){\boldsymbol\Gamma}_{63}\mathbf{u}_3\\
\langle S_z\rangle_{p}^{(i)} &=& \tfrac{1}{4}\,{\boldsymbol\psi_3}^{\dagger}\,\mathbf{J}\,{\boldsymbol\psi_3} = \tfrac{1}{4}\,\mathbf{u}_3^{\dagger}\,{\boldsymbol\Gamma}_{36}\,\Big(\bm \psi^{(1)}_{ij}\Big)^\dagger\,\mathbf{J}\,\Big(\bm \psi^{(1)}_{ij}\Big){\boldsymbol\Gamma}_{63}\mathbf{u}_3.
\end{eqnarray}

The power flux for the reflected and transmitted fields are:
\begin{eqnarray}
\langle S_z\rangle_{s}^{(r)} &=&  \tfrac{1}{4}\,\big(\mathbf{R}^{t}_{s12} \mathbf{u}_3\big)^\dagger{\boldsymbol\Gamma}_{36}\,\Big(\bm \phi^{(1)}_{ij}\Big)^\dagger\,\mathbf{J}\,\Big(\bm \phi^{(1)}_{ij}\Big){\boldsymbol\Gamma}_{63}\mathbf{R}^{t}_{s12}\mathbf{u}_3\\
\langle S_z\rangle_{p}^{(r)} &=&  \tfrac{1}{4}\,\big(\mathbf{R}^{t}_{p12} \mathbf{u}_3\big)^\dagger\,{\boldsymbol\Gamma}_{36}\,\Big(\bm \psi^{(1)}_{ij}\Big)^\dagger\,\mathbf{J}\,\Big(\bm \psi^{(1)}_{ij}\Big){\boldsymbol\Gamma}_{63}\mathbf{R}^{t}_{p12}\mathbf{u}_3 \\
\langle S_z\rangle_{s}^{(t)} &=& \tfrac{1}{4}\,\big(\mathbf{T}^{t}_{s12} \mathbf{u}_3\big)^\dagger\,{\boldsymbol\Gamma}_{36}\,\Big(\bm \phi^{(1)}_{ij}\Big)^\dagger\,\mathbf{J}\,\Big(\bm \phi^{(1)}_{ij}\Big){\boldsymbol\Gamma}_{63}\mathbf{T}^{t}_{s12}\mathbf{u}_3\\
\langle S_z\rangle_{p}^{(t)} &=&  \tfrac{1}{4}\,\big(\mathbf{T}^{t}_{p12} \mathbf{u}_3\big)^\dagger\,{\boldsymbol\Gamma}_{36}\,\Big(\bm \psi^{(1)}_{ij}\Big)^\dagger\,\mathbf{J}\,\Big(\bm \psi^{(1)}_{ij}\Big){\boldsymbol\Gamma}_{63}\mathbf{T}^{t}_{p12}\mathbf{u}_3.
\end{eqnarray}
where the total layer reflection and transmission matrices are defined via their respective transfer matrices:
\begin{eqnarray}
 \mathbf{M}^t_s &=& \mathbf{M}^{-1}_{\rm interface}\,\mathbf{M}_p\,\mathbf{M}_{\rm interface} \\
    \mathbf{M}^t_p &=& \Big(\bm \psi^{(2)}_{ij}\Big)^{-1}\Big(\bm \phi^{(2)}_{ij}\Big) 
    \,\mathbf{M}^t_s\,
    \Big(\bm \phi^{(2)}_{ij}\Big)^{-1}\Big(\bm \psi^{(2)}_{ij}\Big) \\
    \mathbf{T}^{t}_{s12} &=& \mathbf{T}_{12}\big(\mathbf{M}^t_s\big)\\
    \mathbf{R}^{t}_{s12} &=& \mathbf{R}_{12}\big(\mathbf{M}^t_s\big)\\
    \mathbf{T}^{t}_{p12} &=& \mathbf{T}_{12}\big(\mathbf{M}^t_p\big)\\
    \mathbf{R}^{t}_{p12} &=& \mathbf{R}_{12}\big(\mathbf{M}^t_p\big)
\end{eqnarray}
This allows us to define the power imbalance taking into account the directions of power propagation
\begin{eqnarray}
        \Delta_s&=&<S_z>^{(t)}_s-<S_z>^{(r)}_{s}-<S_z>^{(i)}_{s} \; \neq 0 \quad\quad \text{(for evanescent waves)}\\
        \Delta_p&=&<S_z>^{(t)}_p-<S_z>^{(r)}_{p}-<S_z>^{(i)}_{p} \;=0
\end{eqnarray}

In the standard propagation base, we observe an imbalance which is present because of interference between the standard modes when evanescent. In this case, the power flux incident on the layer is dependent on the layer reflectivity. Stated in a different way, the power emitted depends on the properties of a remote layer, making this representation non-local. 

This non-local dependence disappears, and balance is restored when describing the fields using the power mode representation. Incident power and reflected power are orthogonal to each other in this representation and can be treated as distinct. Emitted waves and subsequent reflections can be seen as distinct and independent events and processes in this representation.  

\end{appendices}

\clearpage
\bibliography{ref}

@article{supmat,
  author       = {Michael Mazilu and Andriejus Dem\v{c}enko},
  title        = {Code underpinning - Representation-induced superposition breakdown in linear physics},
  year         = {2026},
  publisher    = {University of St Andrews Research Portal},
  doi          = {10.17630/da4b8bd9-0af6-48f7-bb47-babd40ff038c},
  url          = {https://doi.org/10.17630/da4b8bd9-0af6-48f7-bb47-babd40ff038c}
}

@book{bohren1983absorption,
  author    = {Bohren, Craig F. and Huffman, Donald R.},
  title     = {Absorption and Scattering of Light by Small Particles},
  year      = {1998},
  publisher = {John Wiley \& Sons},
  address={Weinheim},
  doi={10.1002/9783527618156}
}

@article{foldy1945multiple,
  author    = {Foldy, Leslie L.},
  title     = {The multiple scattering of waves. I. General theory of isotropic scattering by randomly distributed scatterers},
  journal   = {Physical Review},
  volume    = {67},
  number    = {3-4},
  pages     = {107--119},
  year      = {1945},
  doi       = {10.1103/PhysRev.67.107}
}

@book{joannopoulos2008photonic,
 author = "Joannopoulos, \{John D.\} and Johnson, \{Steven G.\} and Winn, \{Joshua N.\} and Meade, \{Robert D.\}",
  title     = {Photonic Crystals: Molding the Flow of Light},
  year      = {2011},
  publisher = {Princeton University Press},
  address = "United States",
  doi={10.2307/j.ctvcm4gz9}
}

@article{anderson1958absence,
  author    = {Anderson, Philip W.},
  title     = {Absence of diffusion in certain random lattices},
  journal   = {Physical Review},
  volume    = {109},
  number    = {5},
  pages     = {1492--1505},
  year      = {1958},
  doi       = {10.1103/PhysRev.109.1492}
}

@book{Yeh1988,
  author = {Yeh, Pochi},
  title = {Optical Waves in Layered Media},
  publisher = {Wiley-Interscience},
  year = {2005},
  address = {New York}
}

@book{BornWolf1999,
  author    = {Max Born and Emil Wolf},
  title     = {Principles of Optics: Electromagnetic Theory of Propagation, Interference and Diffraction of Light},
  publisher = {Cambridge University Press},
  address   = {Cambridge},
  year      = {2013},
  doi       = {10.1017/CBO9781139644181}
}

@book{Jackson1999,
  author    = {John David Jackson},
  title     = {Classical Electrodynamics},
  publisher = {Wiley},
  address   = {New York},
  year      = {1999}
}

@book{NovotnyHecht2012,
  author    = {Lukas Novotny and Bert Hecht},
  title     = {Principles of Nano-Optics},
  publisher = {Cambridge University Press},
  address   = {Cambridge},
  year      = {2012},
  doi       = {10.1017/CBO9780511794193}
}

@article{KoSambles1988,
  author  = {D. Y. K. Ko and J. R. Sambles},
  title   = {Scattering matrix method for propagation of radiation in stratified media: attenuated total reflection studies of liquid crystals},
  journal = {Journal of the Optical Society of America A},
  year    = {1988},
  volume  = {5},
  number  = {11},
  pages   = {1863--1866},
  doi     = {10.1364/JOSAA.5.001863}
}

@book{Marcuse1991,
  author    = {Dietrich Marcuse},
  title     = {Theory of Dielectric Optical Waveguides},
  edition   = {2},
  publisher = {Academic Press},
  address   = {Boston},
  year      = {1991},
  doi = {https://doi.org/10.1016/B978-0-12-470951-5.X5001-X}
}

@book{LandauLifshitz1977,
  author    = {L. D. Landau and E. M. Lifshitz},
  title     = {Quantum Mechanics: Non-Relativistic Theory},
  series    = {Course of Theoretical Physics},
  volume    = {3},
  publisher = {Pergamon Press},
  address   = {Oxford},
  year      = {1977},
  doi ={10.1016/C2013-0-02793-4}
}

@article{Berreman1972,
  author  = {Dwight W. Berreman},
  title   = {Optics in Stratified and Anisotropic Media: 4{\texttimes}4-Matrix Formulation},
  journal = {Journal of the Optical Society of America},
  year    = {1972},
  volume  = {62},
  number  = {4},
  pages   = {502--510},
  doi     = {10.1364/JOSA.62.000502}
}

@article{Sprague2024UnitarityQI,
  title = {Unitarity constrains the quantum information metrics for particle interactions},
journal = {Nuclear Physics B},
volume = {1018},
pages = {116989},
year = {2025},
issn = {0550-3213},
doi = {https://doi.org/10.1016/j.nuclphysb.2025.116989},
url = {https://www.sciencedirect.com/science/article/pii/S0550321325001981},
author = {Shanmuka Shivashankara and Hobbes Sprague}
}

@article{Briceno2024PhotonicQC,
  author       = {Ra{\'u}l A. Brice{\~n}o and Robert G. Edwards and Miller Eaton and Carlos Gonz{\'a}lez{-}Arciniegas and Olivier Pfister and George Siopsis},
  title        = {Toward coherent quantum computation of scattering amplitudes with a measurement-based photonic quantum processor},
  journal      = {Physical Review Research},
  volume       = {6},
  pages        = {043065},
  year         = {2024},
  doi          = {10.1103/PhysRevResearch.6.043065},
  url          = {https://link.aps.org/doi/10.1103/PhysRevResearch.6.043065}
}

@article{Kucera2025CurrentSMatrix,
  author       = {Jan Ku{\v c}era and Ulrich Wulf and George Alexandru Nemnes},
  title        = {Scattering Theory in an {N}-Pole Semiconductor Quantum Device: The Unitarity of the Current {S}-Matrix and Current Conservation},
  journal      = {Micromachines},
  volume       = {16},
  number       = {3},
  pages        = {306},
  year         = {2025},
  doi          = {10.3390/mi16030306},
  url          = {https://www.mdpi.com/2072-666X/16/3/306}
}

@article{Wojcik2021,
   title = {Universal Behavior of the Scattering Matrix Near Thresholds in Photonics},
  author = {Wojcik, Charles C. and Wang, Haiwen and Orenstein, Meir and Fan, Shanhui},
  journal = {Phys. Rev. Lett.},
  volume = {127},
  issue = {27},
  pages = {277401},
  numpages = {6},
  year = {2021},
  month = {Dec},
  publisher = {American Physical Society},
  doi = {10.1103/PhysRevLett.127.277401},
  url = {https://link.aps.org/doi/10.1103/PhysRevLett.127.277401}
}

@article{He2024,
  author       = {Huan He and Zhaoxian Chen and Huanan Li and Cheng-Hou Tu and Jingjun Xu and Andrea Al\`u},
  title        = {Evanescent Wave Spectral Singularities in Non-Hermitian Photonics},
  journal      = {Physical Review B},
  volume       = {109},
  number       = {L041405},
  year         = {2024},
  doi          = {10.1103/PhysRevB.109.L041405},
  url          = {https://link.aps.org/accepted/10.1103/PhysRevB.109.L041405}
}

@article{ChouChau2025,
  author       = {Yuan-Fong Chou Chau},
  title        = {Nanophotonic Materials and Devices: Recent Advances and Emerging Applications},
  journal      = {Micromachines},
  volume       = {16},
  number       = {8},
  pages        = {933},
  year         = {2025},
  doi          = {10.3390/mi16080933},
  url          = {https://www.mdpi.com/2072-666X/16/8/933}
}

@article{Zhu2025,
  author       = {Rui Zhu and Chenjiang Qian and Shan Xiao and Jingnan Yang and Sai Yan and Hanqing Liu and Deyan Dai and Hancong Li and Longlong Yang and Xiqing Chen and Yu Yuan and Danjie Dai and Zhanchun Zuo and Haiqiao Ni and Zhichuan Niu and Can Wang and Kuijuan Jin and Qihuang Gong and Xiulai Xu},
  title        = {Full polarization control of photons with evanescent wave coupling in the ultra subwavelength gap of photonic molecules},
  journal      = {Light: Science \& Applications},
  volume       = {14},
  pages        = {114},
  year         = {2025},
  doi          = {10.1038/s41377-025-01794-1}
}

@article{Wang2025,
  author       = {Hui Wang and Timothy C. Ralph and Jelmer J. Renema and Chao-Yang Lu and Jian-Wei Pan},
  title        = {Scalable photonic quantum technologies},
  journal      = {Nature Materials},
  year         = {2025},
  doi          = {10.1038/s41563-025-02306-7}
}

@article{AbuGhanem2026,
  author       = {Muhammad AbuGhanem},
  title        = {Toward scalable fault-tolerant photonic quantum computers},
  journal      = {The Journal of Supercomputing},
  volume       = {82},
  pages        = {51},
  year         = {2026},
  doi          = {10.1007/s11227-025-08132-7}
}

@article{Kolarovszki2025,
  author       = {Zolt{\'a}n Kolarovszki and Tomasz Rybotycki and P{\'e}ter Rakyta and {\'A}goston Kaposi and Boldizs{\'a}r Po{\'o}r and Szabolcs J{\'o}czik and D{\'a}niel T. R. Nagy and Henrik Varga and Kareem H. El-Safty and Gregory Morse and Micha{\l} Oszmaniec and Tam{\'a}s Kozsik and Zolt{\'a}n Zimbor{\'a}s},
  title        = {Piquasso: A photonic quantum computing simulation software platform},
  journal      = {Quantum},
  volume       = {9},
  pages        = {1708},
  year         = {2025},
  doi          = {10.22331/q-2025-04-15-1708},
  url          = {https://quantum-journal.org/papers/q-2025-04-15-1708/}
}

@article{Nellambakam2023,
  author       = {Yuganand Nellambakam and
                  K. Haritha and
                  K.~V.~S. Shiv Chaitanya},
  title        = {Evanescent wave in multiple slit diffraction and n-array antennas in metamaterial using {Cesàro} convergence},
  journal      = {Scientific Reports},
  year         = {2023},
  month        = jun,
  volume       = {13},
  number       = {1},
  pages        = {9981},
  doi          = {10.1038/s41598-023-36894-8}  
}

@article{Takatsu.Sebilleau.2022, 
year = {2022}, 
title = {{Simple renormalization schemes for multiple scattering series expansions}}, 
author = {Takatsu, Aika and Tricot, Sylvain and Schieffer, Philippe and Dunseath, Kevin and Terao-Dunseath, Mariko and Hatada, Keisuke and Sébilleau, Didier}, 
journal = {Physical Chemistry Chemical Physics}, 
issn = {1463-9076}, 
doi = {10.1039/d1cp05530e}, 
pmid = {35188153}, 
pages = {5658--5668}, 
number = {9}, 
volume = {24}
}

\end{document}